# A complex network perspective on brain disease

David Papo[1,2,*], J.M. Buldú[3]

[1] Department of Neuroscience and Rehabilitation, Section of Physiology, University of Ferrara, Ferrara (Italy)
[2] Center for Translational Neurophysiology, Fondazione Istituto Italiano di Tecnologia, Ferrara (Italy)
[3] Complex Systems Group & G.I.S.C., Universidad Rey Juan Carlos, Madrid (Spain)
* Email address: david.papo@unife.it

ABSTRACT

If brain anatomy and dynamics have a genuine complex network structure as it has become standard to posit, it is also reasonable to assume that such a structure should play a key role not only in brain function but also in brain dysfunction. However, exactly how network structure is implicated in brain damage and whether at least some pathologies can be thought of as 'network diseases' is not entirely clear. Here we discuss ways in which a complex network representation can help characterising brain pathology, but also subjects' vulnerability to and likelihood of recovery from disease. We show how the way disease is defined is related to the way function is defined and this, in turn, determines which network property may be functionally relevant to brain disease. Thus, addressing brain disease 'networkness' may shed light not only on brain pathology, with potential clinical implications, but also on functional brain activity, and what is functional in it.

CONTENTS

## 1. Introduction

Network neuroscience characterises brain anatomy, dynamics and ultimately function by endowing them with a network representation and describing them in terms of properties of this representation (Bullmore and Sporns, 2009). A network is a particular structure, i.e. a set of objects with some additional feature on the set, whose objects are nodes and whose additional feature is a relation among the nodes which defines links relating them (Boccaletti et al., 2006). All the information in a networked system is encoded in the relational pattern, whose properties allow not only describing the system's relational structure at all scales but also, at least in principle, explaining properties such as its efficiency and robustness (Papo et al., 2014), and ultimately conceiving of ways it can be steered towards desirable states or away from undesirable ones (Liu and Barabási, 2016; Papo, 2019b).

Whether the brain actually behaves as a complex network or else such a structure merely constitutes a convenient representation is a fundamental yet still poorly understood question (Papo and Buldú, 2024). Showing that at least some brain diseases, which we shall call *network disease*s, are somehow related to brain network structure would constitute an important step towards a proof of genuine brain *networkness*.

Characterising the nature of disease, identifying ways in which it may become observable, even in the absence of behavioural symptoms, quantifying its severity, characterising, and predicting the way it starts and evolves and ultimately finding ways in which it can be acted upon are fundamental theoretical and clinical questions. In the remainder, we evaluate how a network representation may help addressing these questions and defining and understanding brain disease. Importantly, we do not aim at characterising the network structure of any given pathology, but rather at evaluating the way network structure can help addressing generic issues related to brain pathology, from the phenomenology of disease to subjects' vulnerability to or ability to recover from it.

To delineate the role of network structure in brain disease, we first discuss possible meanings of disease, we then briefly review ways in which the brain can be endowed with a complex network structure. Finally, we analyse how complex networks can be used to make sense of various aspects of brain pathology and how various neurological and psychiatric pathologies can be understood in terms of such a structure.

Throughout, we differentiate between two pairs of terms often somehow confusingly used as synonyms in the network neuroscience literature: structure and anatomy, and dynamics and function. We use the term *structure*, which equally applies to anatomy and dynamics, to designate a set of entities and their relations. In particular, we prefer the expression *anatomical* to *structural network*, as it more clearly designates anatomy equipped with a particular structure encoded in a graph representation. On the other hand, as we further explain below, we consider genuine *functional brain activity* to be distinct from bare *brain dynamics* in that it is the particular aspect of brain activity which is used to accomplish a given function.

## 2. Defining disease

*2.1. The semantic field of disease*

There are a number of conceptually different ways to understand disease. This variety is given away by the semantic field of disease and particularly by the etymology of words thought of as its synonyms (Lalande, 1927; Canguilhem, 1943). As in the term *disease* (lack of ease or comfort), *pathology* (from the Greek *πάθος*, suffering) and *ailment* (which causes to ail, to suffer) are referred to effects, particularly to the associated pain. From the same legal semantic field, the terms *damage* (from the Latin *damnum*), *lesion* (from the Latin *laedo*, to injur) and *injury* (from the Latin *iuris*, law, the prefix *in-* expressing negation) refer to the consequences of a violation. The term *abnormal*, initially legal in nature, also refers to a norm violation. Originally expressing a judgement of value rather than a description (Canguilhem, 1943), in the context of disease, it implicitly refers to a different (i.e. healthy) state. Finally, while the terms *disorder* and *anomaly* (from the Greek *ομαλός*, smooth, rather then from *νόμος,* law) refer to structural properties, in mathematical terms respectively symmetry and regularity/differentiability, an explicitly functional aspect underlies the term *malady* (from the Latin *malaptus*). Finally, the term *condition* (from the Latin verb *condo*, to put together, to build), often resorted to in order to avoid unwanted associations induced by these various etymologies, designates an identifiable state resulting from the assembly of various parts or the way this comes into being.

It is also interesting to look at the semantic field of the proneness to disease. The term *vulnerability* comes from the Latin word *vulnus*, injury, itself coming from the Latin verb *vello*, to tear or to lacerate, while *fragility* comes from the Latin verb *frango*, to break. The adjective *critical* and the corresponding noun *crisis* (from the Greek noun *κρίσις* and the verb *κρίνειν*) originally mainly designated separation (of wheat grain from straw and chaff), before acquiring by translation the meaning of choosing (Chantraine, 1968). Thus, *crisis* designates a choice made after a separation, hence perhaps the associated negative meaning stemming from the need to choose to separate something from something else.

Correspondingly, there is more than one way of conceiving of brain disease from a system-level perspective: one consists in characterising its phenomenology in terms of its consequences on healthy brain anatomy and dynamics. In this sense, characterising brain pathology entails defining healthy brain function, which can be represented as a structure or as a field, and pathology some perturbation of it, respectively with a mass (e.g. a mass-invading tumour) or as a dynamical system, in the simplest case, an impulse. A somehow dual way involves defining disease as an emergent property of brain dynamics under certain conditions, e.g. within given ranges of control parameter values. Brain disease can also be seen as a process evolving on an underlying structure, with which it may or may not interact. This can for instance take the form of a generic spreading process, e.g. a transport process (of a diffusive kind if what is modelled is activity propagation, or of an advective one if mass is transported).

*2.1.1. Brain-disease relationships*

These approaches implicitly suggest various possible relationships between brain and disease. In an ontological approach, brain and disease may constitute separate entities, each with its own mass and corresponding spatial localisation. If these two entities interact, disease can be seen as a perturbation of an otherwise intact structure. In a different approach, disease is a particular state of the system. Such an approach can be seen to trace back to Hippocrates' dynamic conceptualization of disease, with exogenous factors only playing the role of context rather than cause (Canguilhem, 1943). In more modern terms, disease may arise when a physiological system operates in a range of control parameters that induces a dynamic regime associated with functional consequences deemed pathological (Mackey and Glass, 1977; Glass and Mackey, 1979; Glass and Mackey, 1988; Glass, 2015). In this sense, pathology is an aspect of the brain's repertoire, which is expressed under certain conditions (McIntosh and Jirsa, 2019).

The dynamic understanding of disease can be approached through two complementary angles, respectively related to the nature of the regime itself and to its *pathogenesis* i.e. the mechanisms through which it develops, progresses, and either persists or is resolved. These approaches give rise to some fundamental questions. On the one hand, when and how can disease be described in dynamical terms? Is the pathological regime qualitatively different from the one associated with healthy brain functioning? If disease is associated with qualitative functional changes, then it can be thought of as forcing symmetry breaking (in Hippocrates' approach, it breaks equilibrium between elements). For instance, in Alzheimer's disease (AD), the loss of dynamical connections and the increased randomness in network structure are key indicators of impaired brain function (Tijms et al., 2013). On the other hand, several mechanisms can potentially underlie pathogenesis. When thinking of disease as equilibrium breakdown, the most obvious of such mechanisms is *homeostastis*, i.e. the self-regulating processes ensuring the maintenance of the conditions that are best for the system's operation. When the brain deviates from this condition, as in the case of epilepsy (van Diessen et al., 2013), dysfunction arises. Furthermore, pathology can be thought of as a change in the system's response function. For instance, the *allostasis* (stability through change) model defines health as optimal responsiveness to fluctuations in demand and disease as shrinkage of adaptive variation (Sterling, 2020). More generally, disease may be associated with loss of complexity rather than regularity (West, 2010).

*2.2. Disease dynamics*

If the brain is thought of as a dynamical system, its evolution to disease can be thought of as a process. The disease process can be temporally segmented, e.g. identifying a healthy state (or a state in which the disease is under control, in an incubation or a chronic period) and a pathological state, separated by a pre-pathological state (Scheffer et al., 2024). The dynamics underlying each of these states can be associated with a complex space, possibly with well-identifiable sub-spaces induced by the generic metastable dynamics of the brain (Roberts et al., 2019; Recanatesi et al., 2022). The disease process is ultimately characterised by the way the dynamics tends to flow towards, around or away from a given state and by the corresponding *basin of attraction*, i.e. the set of initial conditions that lead to that particular state under the dynamics.

Characterising neurological and psychiatric disorders and their trajectory requires understanding how the dynamics moves the system from one state to the other under the action of external drives, e.g. how smoothly or abruptly the system may change, how much the system resists such changes at any given state. The following substantial questions arise naturally: how prone or vulnerable is a given system to damage? How easy is it to push it away from a pathological state? How does it revert to its prior configuration after a perturbation? Perhaps more fundamentally, how much damage is needed to compromise function?

*2.3. Disease, brain function, and functional space*

A common defining factor of disease is that it is associated with functional impairment, often accompanied by behaviourally observable consequences. Impairment may range from decreased efficiency to functional breakdown. This corresponds to Siegerist's (1932) understanding of pathology as *physiology with obstacles*. In this approach, pathology can be derived from normal function, from which it is separated by an alien element which complexifies but does not qualitatively alter it (Canguilhem, 1943). In terms of structure, disease can then be seen as an *obstruction* to some part of the functional space.

Defining brain function and what is functional in brain activity are two only seemingly easy tasks. While brain dynamics and function are often used interchangeably, functional brain activity can be thought of as a particular aspect of brain dynamics, corresponding to the set of all neurophysiological configurations that generate a given function.

Genuinely functional activity results from a complex relation between two structures: the structure of the neurophysiological space together with its equivalence classes (possibly made observable by some recording technique) and the one of an abstract space of functions (made observable by some performance measure). Subdivisions in one space, corresponding to equivalence classes with identical properties, are used to define subdivisions in the other (Papo, 2019a; Papo and Buldú, 2024b) (see Figure 1).

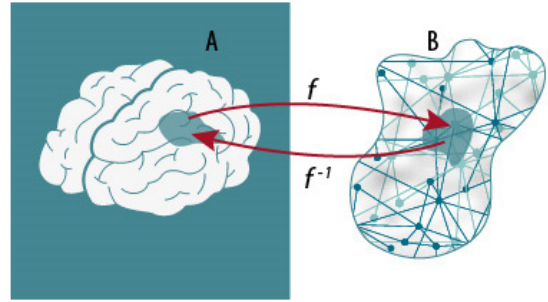

**Fig. 1.** Functional brain activity should not be equated with bare brain dynamics. Functional brain activity results from a complex projections $f$ and its inverse $f^{-1}$ from the structure $\mathcal{S}_A$ of the neurophysiological space **A** to the structure $\mathcal{S}_B$ defined on the abstract space **B** of cognitive functions made observable by some set of performance measures. Ultimately, parcellation in one space is used to define parcellations in the other (Korhonen et al., 2021).

Nearness and neighbourhood relations may qualitatively differ when considering *function* rather than bare *dynamics*. The very definition of other properties such as path-dependence and robustness (Lesne, 2008) may also vary in the dynamics-to-function transition (Papo, 2019a). All these factors may give rise to a space with non-trivial structure (Stadler et al., 2001). Phenomenology including intermittency, redundancy, degeneracy, whereby structurally dissimilar components can perform similar functions under certain conditions but different ones under others (Edelman and Gally, 2001), and functional robustness can be understood in terms of the combination of *accessibility* in the neurophysiological space, i.e. which transitions are realisable in the neighbourhood of underlying neuronal configurations, and the presence of large neutral regions in the space of system configurations or parameters that give rise to equivalent phenotypic behaviour, so that changes in such spaces have no consequence on the functional space onto which they are projected (Stadler et al., 2001; Stadler and Stadler, 2006; Doyle and Csete, 2011; Papo, 2019a).

*2.4. Vulnerability and resilience*

The impact of disease on brain anatomy, dynamics and function depends on the system's ability to resist to external fields or boundary elements acting on the brain at various scales. Various diseases result from the balance of interactions between the system and the environment, mutations, and homeostatic mechanisms keeping the system within physiological regimes (Nijhout et al., 2014). Brain disorder dynamics is often explained as the result of the interaction between a predisposing vulnerability (or *diathesis*) (e.g. genetic background, predisposing conditions, age), and a perturbation (or *stress*) caused by exogenous factors (e.g. psychological or physical trauma), while the risk of neurological and psychiatric disorders is also related the *dynamic resilience* of the healthy state (Scheffer et al., 2024).

Resilience can be understood in a number of ways (Fisher, 2015; Urruty et al., 2016; Liu et al., 2022; Krakovská et al., 2024). For instance, it may designate the degree to which a system withstands perturbations or errors, resists change and avoids collapsing or being driven towards a qualitatively different stable state (Holling, 1973). For instance, it has been suggested that even far from phase transitions, spontaneous activity's neuronal avalanches may be neutral with respect to perturbations (Martinello et al., 2017). In the context of brain pathology, robustness may quantify the extent to which the brain can withstand damage arising, e.g. from tumours, trauma, or stroke (Farooq et al., 2019). Resilience can also be understood in terms of the system's ability to revert to the original stable state after a perturbation. In this sense, resilience is defined by the relaxation time to the previous stable regime (Holling, 1996), although this notion becomes problematic in an inherently multi- or metastable system. Resilience can also be understood in terms of a system's ability to adapt or transform in response extreme shocks (Holling et al., 2002), and its ability to reorganise around alternative states once pushed away from the original state (Gunderson, 2000). This includes the ability to cope through rapid response, and active recovery strategies. To some approximation, the same conceptual framework can be used to make sense of *disease resilience*, i.e. the tendency of disease to persist.

A system's resilience is defined with respect to a given property, for example some topological network property, or dynamical regime, e.g. balanced state, or some information-related property, e.g. informational capacity (Koetter and Médard, 2003) or, more generally, the ability to perform some prescribed task, e.g. information encoding, and against a given set of perturbations (Kitano, 2007), e.g. node or link perturbation. The system may respond to perturbations by changing some but not all properties in order to preserve some other ones. Furthermore, in the presence of limited resources, optimisation of one function necessarily entails fragility of some other possibly unexpected one (Carlson and Doyle, 1999, 2002).

Dynamic resilience depends on the size and shape of each state's basin of attraction. These differ among individuals and, for a given individual, may change over time. The basins of attraction interact not only with external drives but also with both positive (*autocatalytic*) and negative (*homeostatic*) feedback forces, respectively reinforcing and damping the effects of perturbations. In the presence of weak positive feedback forces (e.g. between insomnia and mood), the system's response to changing conditions tends to be smooth whereas strong enough self-reinforcing autocatalytic feedback can cause the response curve to become folded so that there may be more than one stable state over a range of external conditions and a shift from a healthy to a pathological state may occur abruptly as a tipping point is reached. The distance from the tipping point defines the state's *precariousness* (Walker et al., 2004). Close to a tipping point, the system becomes less resilient and more likely to shift to a qualitatively different state.

For a healthy system, avoiding such points is therefore important. This involves finding early warning signs anticipating their occurrence. Two such signs are represented by *critical slowing down*, a generic dynamical phenomenon occurring in the vicinity of a bifurcation point where the system becomes increasingly slow when recovering from small perturbations back to its original state and fluctuations have characteristic properties viz. long-term memory (Dai et al., 2012, 2013; Liu et al., 2022). In this vein, slowing down of recovery is in general thought of as a generic risk marker (Rikkert et al., 2016), although inherently critical phenomena such as epileptic seizures may show no signs of such slowing down (Wilkat et al., 2019).

## 3. Systems with complex network structure

Hitherto, we have thought of the brain as a generic high-dimensional system with no particular internal structure, i.e. an essentially homogeneous and isotropic system. However, it is reasonable to assume that both brain anatomy and dynamics have some relevant non-random structure in both time and space and that function may somehow be modulated by such a structure.

It has now become standard to think of brain anatomy, dynamics and function as a *complex network* (Bullmore and Sporns, 2009). A complex network is a strongly disordered heterogeneous structure (Dorogovtsev et al., 2008) characterised by non-trivial properties which do not feature in simpler ones, such as lattices or random networks (Albert and Barabási, 2002). Various fundamental questions need to be addressed. What does it mean to be a network? Is such a structure relevant to brain structure, dynamics and function? If so, how does the presence of complex network structure change the picture delineated so far and specifically, what role does network structure play in the way disease affects the brain?

### 3.1. What being a complex network means

Equipping the brain with a network representation has a number of important implications. First, it involves adding structure to the anatomical, dynamical and functional spaces, which can be taken advantage of to quantify brain anatomy and dynamics and ultimately brain function. At the most basic level, this involves *distinguishing* structures. Typical representations isomorphic to the anatomical space become difficult to compare when the physiological space has complex, e.g. non compact geometry due to noise and inter-individual differences. A topological structure does away with metric details and may in principle simplify the task (Stam et al., 2014). Deviations from network structure observed in the healthy brain may themselves have non-trivial network structure (Zanin et al., 2014, 2018). *Proximity* relations in the neurophysiological space can also be defined using perturbation methods (Peters, 2016, 2018). Often, it is also important to quantify how far conditions are from each other both in the space made observable by signs and symptoms and in the corresponding neurophysiological space. The presence of a structure induces distances through which it is in principle possible to quantify various disease-related aspects (Rossi et al., 2015; De Domenico and Biamonte, 2016). For instance, the connectivity matrix in real or phase space can be used as a metric tensor endowing the space with a Riemannian differential manifold structure (Amari and Nagaoka, 2007). The key point is that of translating all these anatomical or dynamical properties including distinguishability, proximity, neighbourhood, accessibility into the way disease affects functionally relevant equivalent properties (Papo, 2019a). If disease implies a network structure modification with respect to the healthy brain, it is useful to identify and quantify such a change, either in a categorical or, ideally, in a quantitative way.

A genuine network structure involves a coarse-graining wherein each portion of some observable space associated with the brain, e.g. the anatomical (real) space or the phase space of the dynamics, is identified with a discrete point, summarising a whole subsystem and the system is described through the relational structure of these subspaces. In analogy with the general way of defining brain function, the networks associated with bare dynamics are better called *dynamical*, while *functional* networks should be reserved for structures inducing partitions of the observable space through measures associated with the ability to perform some well-specified task (Korhonen et al., 2021).

If the brain has genuine complex network structure then disorder is relevant: its presence causes properties to deviate from those of homogeneous systems, and this should have an impact on its dynamics and ultimately on its function. For instance, heterogeneous degree distributions are responsible for novel types of phase transitions (Dorogovtsev et al., 2008). Likewise, we propose that a *network disease* is a condition where non-trivial network structure relevant to brain function is damaged.

A complex network structure comes with some fundamental assumptions. First, network features are statistical properties arising from a great number of microscopic interactions, wherein single degrees of freedom (nodes or links) lose, at least *prima facie*, their identity. In this they profoundly differ from merely connected (usually much smaller) systems, where nodes and links are well identified, each having a specific functional role, and are therefore in no way interchangeable. In such a structure, connectivity serves a purely transport role, whereas in large scale networks it may have a more complex one. Second, complex networks' properties emerge from *connectivity*, not from *collectivity* or *mass action*. Such properties may appear at scales below those of nodes but disappear as a result of coarse-graining. Third, in the appropriate space in which the network is defined, at scales above those of nodes, the system may show genuine *non-locality*, i.e. interaction-induced emergence as opposed to bare anatomical connectivity.

Irrespective of the particular conceptual framework adopted to define disease, these are necessary ingredients not only for brain function's *networkness* (Papo and Buldú, 2024) but also for a genuine network disease.

### 3.2. Brain network structure: space specification

Alongside the way disease and brain function are defined, a third factor determining the role network structure may play in disease is represented by the space in which network structure itself is defined.

Various brain aspects may meaningfully be equipped with such a structure and also, importantly, some may not (e.g. purely feedforward, and ascending or descending systems). The anatomical structure, particularly for spatial scales at which nodes and links can respectively be mapped onto neurons and fibres uniting them, appears to be a particularly obvious candidate. Similar maps can apply to glial cells, proteins, or expressed genes (Thompson et al., 2020). However, other possibilities are also available, from *real space* (anatomy-embedded) dynamics to the *phase space* of brain dynamics and the abstract space associated with pathologies and the relationships among them (see Figure 2). For instance, the relevant network for a given pathology may defined on a gene or a protein space, where nodes are single expressed genes or proteins, and relationships, e.g. interaction between them or co-expression, define links. Note that such a space may or may not have an explicit relationship with the underlying anatomical space. While anatomical networks are often modelled as acting upon a dynamic activity field, of which they constitute the spatial structure, the topology and geometry induced by brain activity merely shadowing such anatomical structure, dynamical networks are often considered in relation to brain *function*, which is typically identified with subspaces of dynamic connectivity's (spatiotemporal) structure.

Another key aspect is represented by the temporal and spatial scales at which a network structure can be defined. At fast time scales, the network structure induced by connectivity is approximately constant. For instance, at macroscopic spatial scales, brain anatomy may be thought of as essentially static at the typical timescales of a neuroimaging experimental session, while when considering local synaptic connectivity, the corresponding timescale may be of the order of milliseconds. However, even the slowest structure, i.e. the one associated with anatomy, can be endowed with its own dynamics at sufficiently long scales, *viz.* at developmental or experimental timescales of the order of seconds and beyond (Papo, 2013b). Considering the brain at developmental or evolutionary as opposed to experimental time scales allows understanding observed anatomy (and dynamical repertoire) as the result of a slow dynamics, the underlying forces of which can be modelled, and used as axes appropriate *morphospaces* onto which particular network instances can be mapped (Corominas-Murtra et al., 2013; Avena-Koenigsberg et al., 2015). Conversely, brain dynamics may be endowed with network structure which can be thought of as static under appropriate conditions. A further important aspect relates to the spatial scales at which genuine non-trivial network structure can be expected.

### 3.3. Structure-dynamics relations

Both at microscopic and at coarse-grained, phenomenological scales the brain can be thought of as a networked dynamical system, where each node is associated with its own phase space of given structure (DeVille and Lerman, 2015). The main question will then be how network structure affects brain dynamics in disease. Note that the presence of a network structure incorporates not only spatial extension as already standard in dynamical disease models, but also a relational structure.

The presence of strong disorder associated with a complex network structure affects various dynamical properties of a system (Boccaletti et al., 2006; Porter and Gleeson, 2016). The brain can be thought of as a heterogeneously driven nonlinear networked system (Sreenivasan et al., 2017). One important question is therefore how the presence of such a structure may affect its response to external fields. Network topology modulates networked systems' response to applied fields and noise (Shinomoto and Kuramoto, 1986; Timme, 2007; DeDeo and Krakauer, 2012; Sonnenschein et al., 2013; Shi et al., 2023). Perturbation–response patterns of sets of given dynamical units may be consistent with classes of network topologies (Shandilya and Timme, 2011; Barzel et al., 2015). More generally, insofar as the presence of disorder affects the system's scaling relations, transport coefficients and response function (Dorogovtsev et al., 2008), a complex network structure can be understood as a state of matter parametrisation, with symmetries intermediate between those of a solid crystal and those of a liquid (Papo, 2013a; Sun et al., 2024).

Network structure may also have profound effects on the system's dynamics. For instance, hierarchical modularity may lead to the emergence of metastability (Caprioglio and Berthouze, 2024). Furthermore, important dynamic phenomenological features require the presence of specific structure. For instance, properties such a multistability and sustained oscillations respectively require positive and negative feedback loops (Thomas, 1981; Zañudo and Albert, 2013). Moreover, in the presence of disorder, the dynamics associated with each node encodes information about its topology (Timme, 2006; Burioni et al., 2014; Pernice et al., 2013; van Meegen et al., 2021).

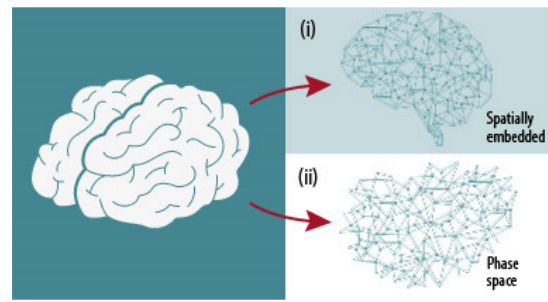

**Figure 2**. Endowing the brain with a network structure means mapping some aspect of its anatomy or dynamics on a set of nodes and links connecting them (Korhonen et al., 2021). This can be done in various ways. For instance, the network may be embedded in the brain's Euclidean space (i), in which case nodes represent brain regions and links either anatomical structure uniting them or some dynamical relation between activity at each node, or in may represent the brain's phase space (ii), in which case nodes may represent brain states and links transitions between them.

Non-trivial network structure also affects processes unfolding on the network (Dorogovtsev et al., 2008) including synchronisation (Boccaletti et al., 2006; Arenas et al., 2008), information transport (Tadić et al., 2007; Gallos et al., 2007; Barzel and Barabási, 2013; Mišić et al., 2015; Gollo et al., 2017; Hens et al., 2019), and scaling properties (Moretti and Muñoz, 2013; Villegas et al., 2014; Millán et al., 2018). For instance, quenched disorder of the anatomical network has been proposed as the structural mechanism through which brain dynamics becomes critical without fine parameter tuning (Moretti and Muñoz, 2013). The interplay between quenched anatomical and intrinsic frequency heterogeneities at various scales may give rise to a regime characterised by large non-trivial spatio-temporal fluctuations (Moretti and Muñoz, 2013) and metastability (Villegas et al., 2014) wherein the order parameter oscillates in a frequency-dependent fashion with the coupling strength and is characterised by well-separated hierarchically organized synchronisation domains of different frequencies (Villegas et al., 2014; Millán et al., 2018).

#### 3.3.1. The role of relative scales

The extent to which such a system's behaviour depends on the network structure on which it unfolds hinges on how dynamics at various spatial scales interact and this, in turn, on the relationship between the time scales at which they evolve. In the simplest case, one may consider the interactions between nodal and network dynamics (Do and Gross, 2012). In this framework, the connectivity matrix may itself be time-dependent and may have its own dynamics and associated set of characteristic scales. If node dynamics is much slower than the network's, what is studied is the evolution of *non-dynamic networks* with static nodes. If, on the other hand, the time scale over which the network evolves is much larger than that of dynamical fluctuations, overall, the system is treated as a static network, and time invariant coupling and network structure heterogeneity acts as *quenched* disorder (Ódor, 2008; Maslennikov and Nekorkin, 2017). When disorder has its own dynamics, network structure is considered to act as *annealed* disorder (Dorogovtsev et al., 2008). Thus, for instance, the anatomical network structure is often thought of as quenched at experimental time scales, whereas at longer (e.g. developmental or evolutionary) time scales, it can no longer be considered to be static and acts as annealed disorder. In addition to local dynamics, one could also consider the dynamics of its control parameters and more generally that of structure at all scales. In such a context, there may no longer be scale separation wherein a slow variable is only affected by the averaged state of the fast variables and the dynamical interplay between the time scales is relatively weak and, at various scales dynamics may potentially interact, a hallmark of *adaptive systems* (Gross and Blasius, 2008; Do and Gross, 2012; Maslennikov and Nekorkin, 2017; Berner et al., 2020).

Furthermore, the timescales of dynamical processes taking place on a network typically differ from both those of the nodes and of the network topology. Thus, in addition to the network structure, the system has two, typically separated scales, *viz.* the slow one of the processes, and the fast one of nodes, each possibly being multiscale. The network structure on which processes unfold is often the anatomical one in the quasi-static limit (Cabral et al., 2011, 2022), but it could also conceivably be the one induced by brain activity. Local dynamics, network structure, and dynamical process all have characteristic times of their own, so that ultimately macroscopic properties may result in a non-trivial way from their mixing (Perra et al., 2012; Lambiotte et al., 2019; Papo et al., 2017). As we discuss in Section 3.6., network neuroscience has been employed as a tool to classify brain diseases based on alterations in specific network parameters, some of which are shared across multiple disorders, while others are unique to particular conditions.

*3.4. From the network structure of function to the function of network structure*

It is natural to think of observed network structure as functionally meaningful (Milo et al., 2002; Kashtan et al., 2005; Ashwin et al., 2024; Bashan et al., 2012), and to assume that a particular wiring pattern determines the system's dynamics and function (Mastrogiuseppe and Ostojic, 2018; Ocker et al., 2017). For instance, it is intuitive to see that transport properties should depend on network structure at all scales (Mišić et al., 2015; Avena-Koenigsberger et al., 2017; Villegas et al., 2022). The brain's computational properties may also crucially depend on network structure. For instance, memory encoding should depend on some property of the network supporting it (Susman et al., 2019). Moreover, network structure can differ vastly in a function-dependent way, and be designed to optimise a given function, e.g. information spreading, but not information processing or storage (Zylberberg et al., 2017).

If brain network properties are related to function in the healthy brain and altered in disease, dual questions are how network structure may enable the brain to carry out such function and bound it in the absence of disease (Petri et al., 2021), and how it may obstruct it in its presence. For instance, the brain can use the topological structure of neural networks in real space to perform computation (Jazayeri and Ostojic, 2021; Ostojic and Fusi, 2024; Ashwin et al., 2024), and such a structure may also explain the system's computational properties and its limits (Curto, 2017).

Following this line of reasoning, network-related equivalence classes would also define functional equivalence classes. However, topological, geometric or combinatorial equivalence does not necessarily entail functional equivalence and whether such equivalence classes underlie brain function is unknown. In other words, these classes do not necessarily coincide with the structural aspects enabling the system to carry out its assigned function. For instance, whether isotopic networks are dynamically and functionally equivalent and, conversely, non-isotopic networks inequivalent is still poorly understood.

A further important aspect relates to scale-dependence of network structure's functional significance. For instance, motifs (Milo et al., 2002) may have an information processing role at mesoscopic but not at larger scales: at system level, network structure may instead support transport rather than computation. Correspondingly, network structure's role should also be scale-dependent in pathology.

More generally, that a system has a given structure does not entail that such a structure is functional. Brain structures are organised to carry out different functions and therefore to satisfy multiple constraints. Different networks can give rise to similar function (Ganmor et al., 2015) while the same network can function differently depending on the context, both distinguishing features of degenerate systems (Edelman, 1987; Hennig, 2023). For instance, the same network motif can perform many different functions depending on functional requirements. Thus, on the one hand, function cannot always univocally be inferred from observed structure and, on the other hand, it is not straightforward to predict structure from function in neural networks (Biswas and Fitzgerald, 2022). Finally, identifying functional classes is not equivalent to identifying the structural aspects enabling the system to carry out its assigned function.

*3.5. Network properties, invariants and equivalence classes*

A complex network has combinatorial (Bollobás, 1986; Erickson, 2014), topological (Boccaletti et al., 2006), and geometric properties (Boguñá et al., 2021), as well as symmetries (Garlaschelli et al., 2010; Pecora et al., 2014; Dobson et al., 2022) which are encoded in the connectivity matrix. Understanding which of these properties turns out to be functionally relevant is central to the network neuroscience effort.

Network neuroscience often focuses on topology, both for real and phase space networks. Topology deals with properties that are invariant under continuous deformation without tearing or cutting, e.g. bending, twisting, and stretching, but does not consider other properties such as size or orientation. In such a framework, nodes and links are dimensionless, node location is disregarded, distances between nodes are not metric but defined as the minimal number of links in the connecting path, and brain networks' structural characteristics are uniquely determined by the adjacency matrix. However, network layout and metric aspects may both play an important role in the brain (Wang and Kennedy, 2016; Henderson and Robinson, 2011, 2013, 2014; Roberts et al., 2016). For instance, the anatomical wiring cannot entirely be accounted for in terms of topological wiring rules and appears to at least partly stem from spatial embedding (Henderson and Robinson, 2011, 2013, 2014; Roberts et al., 2016). Physical wires cannot cross, and this imposes limitations on the system's structure (Bernal and Mason, 1960; Song et al., 2008; Cohen and Havlin, 2010; Dehmamy et al., 2018; Liu et al., 2021), and mechanical properties qualitatively change with node and link density (Dehmamy et al., 2018). Moreover, the system's structure is determined not only by the connectivity matrix, but also by the network's spatial layout (Cohen and Havlin, 2010). Spatially embedded networks may have identical wiring but different geometrical layouts. Indeed, for a given adjacency matrix, a network can have an infinite number of layouts, differing in node spatial positions and wiring geometry. Likewise, topologically equivalent systems may be combinatorially different (Babson et al., 2006).

Network ensembles with given properties constitute equivalence classes. For instance, the layouts with identical wiring that can be mapped into one another without link crossing or cutting define *isotopy classes* (Liu et al., 2021). This in turn allows defining equivalence relations between different systems, and rules to relate different points within or between systems, e.g. neighbourhood or proximity relations. On the other hand, the relationship the various aspects of network structure, e.g. the relationship between topology and geometry in brain networks is still poorly understood.

*3.6. A network-based disease taxonomy*

Brain pathologies differ along numerous characteristics, including aetiology, affected part of the system, associated cognitive dysfunction and behavioural correlates. Correspondingly, diseases differ in dynamics, spatial and temporal scales, predictability, and reversibility. But how does this translate into different roles of network structure implication in disease? Can we categorise brain diseases based on their underlying network structure and their role in pathology?

Considering networks in real space, disease may act on nodes, e.g. as the result of tumours or strokes, or by targeting connections, e.g. via demyelination and axonal injury. In either case, brain disease can induce both localised or widespread functional damage, either by propagating along neural connections to other areas, or by leading to anomalous connectivity (Raj and Powell, 2018). While various neurological and psychiatric disorders including epilepsy (Spencer, 2002; Lehnertz et al., 2009, 2017, 2023; Kramer and Cash, 2012; Richardson, 2012), stroke (Wang et al., 2010; Kuceyeski et al., 2014), traumatic brain injury (TBI) (Pandya et al., 2017), multiple sclerosis (Rocca et al., 2016), autism spectrum disorder (ASD) (Oberman et al., 2014), schizophrenia (van den Heuvel et al., 2010; Fornito et al., 2012), and neurodegenerative diseases including AD (Lo et al., 2010; Buldú et al., 2011; de Haan et al., 2012), frontotemporal dementia (Kuceyeski et al., 2012; Sedeño et al., 2017), and amyotrophic lateral sclerosis (Verstraete et al., 2012) are associated with changes in both global and local network organization, from a system-level perspective, one may distinguish non-local disease from local disease with non-local consequences. For instance, it is straightforward to think of traumatic brain injury (TBI) as a non-local disease. Indeed, TBI typically causes diffuse axonal injury (Adams et al., 1980, 1989; Gentry, 1994; Sidaros et al., 2008; Hellyer et al., 2013), and disruption of long-range fibers (Hulkower et al., 2013). These in turn promote secondary biochemical cascades hours to days after the initial injury (Greer et al., 2011), perturbing axonal transport processes, and ultimately resulting in the accumulation of products and alterations in neuronal homeostasis (Johnson et al., 2013). On the other hand, Parkinson's disease (PD) appears as a focal disease with widespread motor and cognitive consequences. In PD, classic motor features are associated with the progressive neurodegeneration of nigrostriatal dopaminergic neurons in the substantia nigra pars compacta, or in some specific cortical region, as indicated by the focal onset of motor symptoms (Foffani and Obeso, 2018). Intermediate cases of various kinds are represented by schizophrenia, epilepsy, and AD. Schizophrenia has classically been characterised as a disconnection syndrome (Friston, 1998; Stephan et al., 2009; Schmidt et al., 2013; Hahamy et al., 2015; Vasa et al., 2016; Hilary and Grafman, 2017), but non-trivial structure beyond connectivity has also been reported (van den Heuvel et al., 2010, 2013). Disease onset in AD may be relatively focal in the anatomical space, while its progression associated with characteristic functional impairment may be progressively more network-like.

On the other hand, both focal and generalized seizures arise from the dynamics of a distributed, large-scale aberrant network spanning lobes and hemispheres and seizure proneness in any part of the network can be affected by activity far away in the network, and electrical and behavioural phenomenology associated with seizures reflects as the network as a whole (Spencer, 2002). Moreover, while some relationship between spatial–temporal seizure dynamics and global topological properties of the associated dynamical networks has been reported (Schindler et al., 2008; Kramer et al., 2008; Bialonski and Lehnertz, 2013; Burns et al., 2014), local network properties such as node centrality are not predictive of the seizure onset zone (Geier et al., 2015). Thus, epileptogenesis and ictogenesis can be thought of as an emergent property of aberrant large-scale neuronal connectivity, and seizures as a network phenomenon (Lehnertz et al., 2009, 2017, 2023; Kramer and Cash, 2012; Richardson, 2012). Interestingly, while epilepsy appears to be the prototype of diseases emerging as a result of global interactions, it also illustrates the possibility of local dynamic abnormality emerging from global abnormality, as accumulating evidence suggests that even focal epilepsy is associated with diffuse structural and dynamical abnormalities (Engel et al., 2013; Stam, 2014).

Diseases may also differ according to whether they directly affect the anatomical (e.g. AD, schizophrenia) or the dynamical (e.g. epilepsy) network structure, or a combination of both (e.g. AD and epilepsy). But this distinction may not always be clear-cut. For instance, while seizure-prone structures tend to be anatomically connected to damaged regions (Riederer et al., 2008; Bernhardt et al., 2011, 2015; Raj et al., 2010; Chiang and Haneef, 2014), they are in essence of a dynamical nature.

A further way to classify brain diseases is induced by the relationship between anatomy, dynamics, and function. For instance, in epilepsy, the functional space is identified with the dynamical one, as the former becomes trivial during seizures. This differs from other psychiatric diseases, in which dynamics and function are qualitatively different spaces, and non-trivial mapping between them needs to be defined. Finally, the role of network structure may differ across diseases: in some cases, e.g. in tumours or in epilepsy, it may be a structure emerging from disease, while in other cases the network structure's role is that of pushing brain dynamics to pathological regimes.

## 4. Disease as healthy network structure alteration

Whether induced by internal dysfunction or by external factors such as pathogens or trauma, disease may be associated with damage to an entire system or to some of its parts. The notion that stable groups of symptoms are associated with lesions of well defined anatomical regions dates back to Morgagni (1682-1771) and underlies the foundations of *anatomical pathology*.

Equipping the brain with a network structure adds three aspects to this foundational framework. First, if some network property has functional meaning, it is natural to suppose that its modifications should be associated with pathology, and conversely that at least some neurological and psychiatric pathologies should be associated with abnormalities in network structure. Second, the consequences of damage are non-local, i.e. the structure-function map should be complex, irrespective of whether the lesion is itself local or not. Third, brain dysfunction may be thought of as a distance from a functional structure and quantified in terms of the magnitude of such distance.

Two fundamental questions should be addressed. Does disease affect the network structure and if so, what aspects of brain physiology (e.g. neural population or vascular system) and network structure are affected by disease? How *can* disease affect network structure?

*4.1. Connectome alterations in brain pathology*

Both in traumatic brain injury and in neurodegenerative diseases, it is natural to think that disease acts on the brain's network structure, possibly altering its geometry and topology. Brain injury is characterised by *primary damage* which induces immediate degeneration and cell death in in regions proximal to the insult, but also *secondary damage* triggering downstream events which can extend to remote regions spared by primary damage (Park et al., 2004). Secondary damage is typically proportional to primary damage's extension and duration (Block et al., 2005) but of much longer duration. Remote damage can result from axonal damage or from transneuronal degeneration, which can involve neurons losing their projection target or their input (Viscomi and Molinari, 2014). Axonal damage gives rise to a number of local and nonlocal phenomena including Wallerian and Wallerian-like degeneration, ultimately resulting in disconnection and loss of signalling (Seeley et al., 2009), which can conceivably translate into changes in brain network geometry and topology.

Charting the brain overall relational structure is the main goal of recent collective connectome projects (Alivisatos et al., 2012, 2013; Larson-Prior et al., 2013; Van Essen et al., 2012, 2013; Allen et al., 2014; Okano et al., 2015; Glasser et al., 2016; Amunts et al., 2016; Miller et al., 2016; Poo et al., 2016; Howell et al., 2017), as well as of functional cartography (Mattar et al., 2015; Jirsa et al., 2010) and discovery science (Biswal et al., 2010). A similar representation can be associated with gene (Bota et al., 2003; Patania et al., 2019) or neurotransmitter expression (Hansen et al., 2022). While the connectome represents static connectivity between brain areas, the *chronnectome* involves a temporal dimension, describing brain dynamics as a set of reoccurring, temporal connectivity patterns (Calhoun et al., 2014; Iraji et al., 2019).

In the same way that a *connectome* is intended as a complete anatomical, dynamical and even genuinely functional network structure of the healthy brain, *pathoconnectomics* should constitute a map of abnormal brain networks (Rubinov and Bullmore, 2013; Deco and Kringelbach, 2014), the main underlying idea being that the connectome could constitute a biomarker for disease (Kaiser, 2013), and therefore that pathoconnectomics is in fact genuine functional structure.

Already at a purely dynamical level, various fundamental reasons render the connectome induced by dynamics, usually termed *functional connectome*, inherently more complicated to define. First, defining links in dynamical networks is far less straightforward than in the anatomical case as for instance no connectivity metric is explicitly based on neurophysiology. Each of the available metrics is predicated upon assumptions on the way brain units are related to each other and, even more fundamentally, on the functional meaning of such relation. Second, the possibility of such interactions to occur on multiple temporal scales places an additional challenge on the detection of interactions and the timescales of the associated phenomena can be rather non-trivial (Papo, 2013b, 2015). Mapping the dynamical space into the functional one adds a further complexity layer, important reasons being the presence of *neutral networks* in the physiological space and highly degenerate circuits. In a degenerate system, lesions are not necessarily associated with functional deficit, and this makes it difficult to establish whether a brain region is necessary for a given cognitive process (Price and Friston, 2002). To determine whether network structure damage underlies the functional disruption associated with disease it is therefore necessary, first to identify what aspect of brain structure is functionally relevant and then to understand the functional meaning of structure from which the structure-function relationship can ultimately be deduced.

*4.2. Structural consequences of disease*

Connectomes are *per se* bare (anatomical or dynamical) connectivity charts. The role of *connectivity* in brain function (Kozma and Freeman, 2016) and the impact of its increase or decrease in neurological and psychiatric conditions (Geschwind, 1965; Friston, 1998; Geschwind and Levitt, 2007; Casanova and Trippe, 2009; Stephan et al., 2009; Terry et al., 2012; Schmidt et al., 2013; Oberman et al., 2014; Hahamy et al., 2015; Vasa et al., 2016; Hilary and Grafman, 2017) has by now convincingly been documented.

Evidence for *network diseases* in analogy with *dysconnection syndromes*[1] (Geschwind, 1965; Weinberger, 1993; Friston and Frith, 1995; Friston, 1998; Stephan et al., 2009; Dineen et al., 2009; Drzezga et al., 2011; Friston et al., 2016) would go a long way into demonstrating the genuine network-like nature of brain function. Arguably the first step in this direction consists in documenting changes in non-trivial anatomical and dynamical network structure in neurological and psychiatric conditions (see Stam, 2014 and Buldú et al., 2024 for reviews on the topic).

One important question is whether the functional impairment associated with dysconnection syndromes, e.g. AD or schizophrenia, is also associated with changes in non-trivial higher-order network structure induced by variations in connectivity. For instance, alterations of brain dynamics' steady-state structure in AD have been shown to induce loss of non-trivial network structure and a shift towards increasingly random structure (Buldú et al., 2011), consistent with a vision of disease as loss of complexity rather than regularity (West, 2010). On the other hand, schizophrenia has been associated with a reduction in white matter connectivity selectively affecting the so called *rich club*, i.e. the pathways connecting high-degree nodes (van den Heuvel et al., 2010, 2013). Moreover, high-order nonlinear interactions in multi-task networks have been proposed as indicators of aberrant brain function in schizophrenia (Plis et al., 2014). Change within the network may be non-local. For instance, reconfiguration of the epileptic brain network potentially affects any network constituent and their dynamical connections (Lehnertz et al., 2016; Rings et al., 2019b; Lehnertz et al., 2023).

On a somehow related level, a further important question is network structure's specificity and sensitivity to given pathologies. While there may be differences in anatomical and dynamical networks' network properties between a given disease and the healthy brain, this does not *per se* guarantee that the former is a genuine network disease, even when differences between populations, e.g. between fronto-temporal dementia and healthy controls (Sedeño et al., 2017) are only found in terms of network properties.

While the emphasis of connectomics is in general on brain networks' bare connectivity and topology, brain network combinatorics and geometry can also be affected in disease in both real and phase space. For instance, physicality (Pósfai et al., 2024): and topology-geometry interactions presumably play an important role in pathologies such as tumours. It is in particular important to understand not only what role that such factor plays on the network structure and ultimately on brain dynamics and function, but also when this may start being important. Furthermore, pathology can affect anatomical (Simhal et al., 2020) and dynamical network curvature.

*4.3. From structure of the dynamics to the dynamics of structure*

It has long been recognised that at both neurophysiological and behavioural levels at least some diseases have an inherently dynamical nature and undergo bifurcations, i.e. qualitative changes, for some values of parameters controlling the dynamics (Mackey and Glass 1977; Glass and Mackey 1979; Glass and Mackey, 1988; Glass, 2015). However, what role is played by network structure is still not totally clear.

In essence, a dynamical disease with a network support can display bifurcations either of the dynamics or of some network property, e.g. topological phase transitions. These correspond to two aspects of network structure's relationship to dynamics, i.e. the network structure induced by the spatiotemporal organisation of the dynamics and the space within which lives the dynamics induced by the anatomical network. Each of these aspects sheds light onto some role of network structure in brain disease.

*4.3.1. Dynamical consequences of anatomical structure damage*

Up until now, we reviewed possible disease-induced emergent network structure. However, disease may also involve network-induced changes in dynamics. At the most basic level, insofar as complex network structure can be understood as a state of matter (Papo, 2013a; Sun et al., 2024), disease can change the system's spontaneous fluctuations and generic response to external fields (Demetrius, 2013). Moreover, disease may act by shoving brain dynamics away from criticality. Indeed, global scaling properties have been shown to be altered in various pathologies (Oberman et al., 2014; Bruining et al., 2020; Liang et al., 2024). But what role does the network structure play in this action of disease on dynamics?

---

[1] While various pathologies have originally been thought to arise as a result of *disconnection*, it has more recently been proposed that these are perhaps more generally defined as resulting from *dysconnection*, as disease may also result from increased link strength and density (Friston et al., 2016).

Disease may act on dynamics through its action on the anatomical network structure, which may force criticality without fine parameter tuning in the healthy brain (Moretti and Muñoz, 2013). Non-trivial anatomical network structure and its modifications induced by disease could act as *topological defects* (Mermin, 1979; Ódor, 2008; Nishimori and Ortiz, 2010; Bowick et al., 2021; Shankar et al., 2022). Such a structure may be relevant to a number of pathologies. For instance, tumour formation can be associated with collective flocking motion through a medium filled with defects.

Disease may also act directly on the dynamical structure. At the most basic level, this may occur via aberrant dynamic connectivity (Oberman et al., 2014). Non-trivial scaling in networked systems has been proposed to arise from the balance between excitation and inhibition (van Vreeswijk and Sompolinsky, 1996). Thus, the main pathophysiological mechanism in diseases such as ASD and schizophrenia may be represented by altered intracortical inhibition within neural microcircuitry (Eichler and Meyer, 2008; Yizhar et al., 2011). For instance, reduced inhibition in ASD (Rubenstein and Merzenich, 2003; Markram and Markram, 2010; Vattikuti and Chow, 2010) may result in an increased excitation/inhibition ratio (Rubenstein and Merzenich, 2003; Nelson and Valakh, 2015; Dickinson et al., 2016) ultimately modifying the system's macroscopic scaling properties (Liang et al., 2024). Brain disease may induce dynamical changes by disrupting non-trivial structural properties at macroscopic scales. For instance, the loss of hierarchical modularity in pathologies such as AD or schizophrenia may impair brain metastability (Caprioglio and Berthouze, 2024).

*4.3.2. Structure dynamics in disease*

If disease is associated with changes in network structure, it is important to understand the nature of such changes. In this sense, the difference between healthy activity and pathology may be comparable to dynamical and functional switching in the healthy brain (Papo and Buldú, 2024, 2025). Likewise, disease is itself often inherently dynamic nature (Torres et al., 2016; Fülöp et al., 2020; Liu et al., 2022), with varying symptoms and underlying neurophysiology, and this may be mirrored by structural dynamics. One fundamental question involves determining whether such network structure changes are quantitative or qualitative. Various scenarios are in principle possible.

First, the transition to pathology may not be associated with any observable topological change, even in the presence of qualitative behavioural changes. For instance, generalised or focal seizures may arise as a consequence of subtle changes in network structure (Terry et al., 2012). A second possibility is that disease is associated with genuine *topological phase transitions* associated with singularities in global network structure properties in the anatomical or dynamical network structure. Such transitions can emerge as the level of noise associated with rewiring is varied (Derényi et al., 2004, 2005; Palla et al., 2004). Metastable-state transitions with topological changes in the minimal-spanning-tree of the network induced by cross-correlations in electrical brain activity have been reported (Bianco et al., 2007). However, whether network topology in the real space acts as a control parameter enforcing brain dynamics bifurcations at some scale is insufficiently understood. A third possibility is that disease is associated with transitions in the phase space of the dynamics. Under some conditions, topological changes undergone by the level subsets of some configuration property of the manifold are related to microcanonical entropy's singularities (Pettini, 2007; Kastner, 2008; Santos et al., 2014; Casetti et al., 2000). In dynamical brain networks, this can occur for dynamic coupling levels emerging as a connectivity threshold is varied (Santos et al., 2019; de Amorim Filho et al., 2022). This topological change is associated with singularities of the *Euler characteristic*, an integer-valued topological invariant describing the system's associated structure (Matsumoto, 2002; Ghrist, 2014). However, while in systems with short-range interactions topological changes in the configurational space constitute a necessary (though not sufficient) condition for phase transitions to occur (Casetti et al., 2000; Franzosi and Pettini, 2004; Pettini, 2007; Kastner, 2008), for systems with long-range interaction potentials, topological properties of the potential energy may not change remarkably at a second-order phase transition (Campa et al., 2009). Moreover, topological changes may be numerous even away from a phase transition (Kastner and Schnetz, 2008). A necessary criterion for the occurrence of a thermodynamic phase transition relates to the curvature at the saddle points of the potential (Kastner and Schnetz, 2008). Disease could also be associated with anomalous scale-dependent topological cross-overs (Rozenfeld et al., 2010; DeDeo and Krakauer, 2012; Villegas et al., 2022). Indeed, healthy brain activity may be characterised by hierarchical organisation into modules with large-world self-similar properties at macroscopic scales, and the addition of only a few weak links can turn the system into a non-fractal small-world network (Gallos et al., 2012), but how this may possibly change as a result of brain pathology hasn't been directly addressed.

A further question requires understanding which of the various aspects of network structure (topology, geometry, combinatorics) may be affected in disease. For instance, disease may affect brain network geometry without affecting topology, as for instance in tumour invasion. Likewise, in damage to memory-related function, e.g. in the hippocampus, combinatorial transitions may induce functional impairment without changing topology.

*4.4. Role of network structure in the functional consequences of disease*

Pathology typically affects the brain's ability to carry out its functions, by reducing its efficiency and reliability (Schwarze et al., 2024), sometimes leading to complete functional breakdown. Disease-related structural changes may lead not only to dynamical and functional inefficiency but also to thermodynamical one (Zhang and Raichle, 2010), although the relationship between network structure and thermodynamics is still poorly understood. But does disease affect brain function by altering its network structure? For instance, does impairment in information spreading, encoding, and processing proceed from structure damage? Can the loss of efficiency in the way the brain carries out the function it is supposed to perform and manages the cost-efficiency trade-offs that it faces be quantified in terms of network properties? Conversely, can functional changes force network structural changes?

*4.4.1. Network scales of disease action*

When considering a neural population in the real anatomically-embedded space, the first question is to do with the network level at which disease may act. Functional damage may occur at scales at which there is no network-like activity; conversely, at scales for which such an activity is present, damage may be renormalized within single nodes. Thus, at the scales at which network structure may be functionally relevant, disease-related perturbations may act not directly upon the network topology, but upon nodal dynamics. For instance, in epileptic seizures macroscale dynamics may be driven by microscale neuronal activity (Burrows et al., 2023). This may change the system's dynamics without modifying anatomical or even dynamical network structure (or function).

While the meaning of nodes and links is in general scale-dependent, it has been proposed that at system level, connectivity may carry information about disease severity, while local node parameters may be related to etiology and prognosis (Zonca et al., 2024).

*4.4.2. Network structure and activity propagation in disease*

One fundamental way in which disease may affect brain function efficiency is by changing the way activity propagates across the space. Indeed, network efficiency is in general framed in terms of propagation rather than of the ability of a given networked system to perform some task. In turn, how activity spreads through a network is often characterised not directly in terms of the propagation process, whose nature is in general unknown, but somehow indirectly in those of the substrate topology visited by such process (Latora and Marchiori, 2001).

How perturbations propagate through networks, impact and disrupt their functions may depend not only on the type and location of the perturbation, but also on the interplay between topological properties, interaction dynamics, and self-dynamics (Avena-Koenigsberger et al., 2018; Hens et al., 2019; Bao et al., 2022).

While the most obvious way in which disease can affect activity propagation is through outright disconnection, network properties ranging from local ones such as node degree, to mesoscale structure such as motifs and hubs can shape global communication and facilitate integrative function (Mišić et al., 2015), determining the scaling regimes and propagation properties (Bao et al., 2022). Importantly, damage may also hinder activity propagation across scales, even in the absence of anatomical damage (Ghavasieh et al., 2023).

*4.4.3. Network structure and information transport and processing in disease*

Brain pathology can induce a range of functional deficits with characteristic behavioural correlates. While these deficits may often be thought of as the result of changes in brain transport properties, they may also emerge in alternative ways.

At the most basic level, it is important to understand the role of network structure in the way brain response to external stimuli may vary as a result of disease. When considering the relationship between network structure and response function of a given system, two dual aspects should be considered. So far, we considered network structure changes associated with given pathologies. However, a second factor is represented by the way network topology affects the response to external fields. The brain responds to external stimuli by computing an implicit model of the environmental variables through transient patterns of dynamics (Tsuda, 2001). As it performs computation the brain faces a trade-off between sensitivity and reliability. On the one hand transient dynamics must be input-specific. On the other hand, computations must be reproducible, robust against noise and variations in initial conditions, and easily decoded. Input-specific yet structurally reliable transient dynamics can be achieved with a heteroclinic dynamics (Rabinovich et al., 2008; Ashwin et al., 2024). In a high-dimensional networked system, this can be achieved with asymmetric inhibitory connections (Nowotny and Rabinovich, 2007). This type of connectivity can coordinate neuronal populations' sequential activity and stabilise heteroclinic channels (Rabinovich et al., 2008). While higher-network structure also plays a role in the response function (Sreenivasan et al., 2017), the role of non-trivial network structure in the response function alterations in disease, including the scale at which this may happen and the properties through which disease can act, is still poorly understood.

One intuitive way to understand the functional role of network structure in disease consists in interpreting structure in energetic and evolutionary terms. For instance, the energetic cost associated with the rich club structure points to a functionally advantageous structure, and rich club hypoconnectivity in schizophrenia likely affects the system's communication capacity. However, it is not straightforward to explain altered brain dynamics and behavioural phenomenology in schizophrenia in these terms. A functionally more interpretable aspect may be represented by the breakdown of modularity (David, 1994; Alexander-Bloch et al., 2010; Godwin et al., 2015). For both anatomical and dynamical networks, a modular structure is characterized in real space by the presence of groups of nodes with dense or strong intrinsic but sparse or weak extrinsic connectivity. Modularity is associated with structure at various, possibly hierarchically organized (Pathak et al., 2024) spatial (Meunier et al., 2010) and corresponding timescales, naturally introducing separation between intra and inter modular scales. Correspondingly, a modular structure may constitute a functionally efficient architecture, provided it realises an optimal balance between local segregated and global integrated activity modes (Tononi et al., 1994; Tononi and Edelman, 2000). Functional impairment associated with dysmodularity may stem from either intra- or inter-module dysconnectivity. In both cases, this can emerge from both hypo- and hyperconnectivity. From the developmental perspective relevant to schizophrenia or ASD, this may respectively correspond to excessive and insufficient cortical pruning. For instance, ASD has been characterised both as a developmental hypo- (Geschwind and Levitt, 2007), and hyperconnection syndrome (Casanova and Trippe, 2009). However, while emerging deficits in schizophrenia may result from excessive cortical pruning in a way similar to what may happen in a neurodegenerative disease, decreased global pruning seems a more plausible model. In the presence of increased anatomical and dynamical connectivity induced by reduced pruning, modules can become more interconnected, with shortcuts among modules allowing changes to propagate. This may explain the impairment of supramodular functions characteristic of schizophrenia, including deficits in perceptual filtering (McGhie and Chapman, 1961), vulnerability to overload, proneness to perceptual interference (Liddle and Morris, 1991), overflow of linguistic analysis into perceptual systems, resulting in thought insertion and auditory hallucinations, tendency to spurious associations (Frith, 1992). The possible presence of hierarchical modularity adds further accessible ways through which pathology may arise from network structure perturbations. For instance, schizophrenia has been shown to be associated with enhanced global hierarchical organization at rest (Acero-Pousa et al., 2024). However, while a structural change is in turn associated with negative symptoms including apathy and formal thought, it is unclear how specifically the former can be used to explain the functional aspect of the latter.

Conversely, disease-related network structure can emerge as the result of functional changes associated with disease. The role of network structure in disease-related information processing impairment can be understood by considering topological network features such as heterogeneity and modularity as emergent properties of functional trade-offs (Zylberberg et al., 2017; Ghavasieh and De Domenico, 2024). For instance, it has been recently proposed that mesoscale structure can emerge from a trade-off between information exchange and response diversity in a wide range of time scales (Ghavasieh and De Domenico, 2024). Thus, disease-related changes in network structure may point to changes in the balance between information transport and processing (Zylberberg et al., 2017; Ghavasieh and De Domenico, 2024).

## 5. Network structure of disease and brain-disease interactions

In some instances, it may be natural to think of disease as a separate system. As consequences, disease itself may have its own structure. Such a structure can be understood as a complex network at three different levels: in the space of diseases, in the behavioural space of observable behavioural symptoms, and in the underlying space of neurophysiological processes from which these symptoms arise.

The idea that the space of pathologies has a certain order, upon which the discipline of *nosography* is predicated, dates back to Thomas Sydenham (1624-1689). It is then natural to chart brain disorders and the relations among them (Cristino et al., 2014; Fornito et al., 2015; Halu et al., 2019; Guloksuz et al., 2017). The space of of human diseases and disorders may be endowed with a network structure induced by genetic or protein interactions (Goh et al., 2007). Moreover, diseases, particularly within a geographical area, may induce a space where single diseases may interact in various ways (e.g. antagonistic or synergistic), a notion known as *pathocenosis* (Nicolle, 1933; Grmek, 1969; Gonzalez et al., 2010). In graph theoretical terms, such a space is naturally thought of as a network-of-networks (Gao et al., 2011; Kenett et al., 2015; Kiani et al., 2021). Within the functional space made observable by clinical phenotypic manifestations, brain disorders can be thought of as systems of subject-specific causally connected symptoms rather than as effects of a latent disorder (Borsboom et al., 2013; Zhou et al., 2014; Guloksuz et al., 2017; Epskamp et al., 2018; Tosi et al., 2024). Diseases share not only signs and symptoms, but also underlying physiology, including genes and proteins, as well as therapies to treat them. Finally, in some cases, for instance involving invading masses and tumours, it is also natural to ascribe to disease a structure of its own and distinguishable from that of the healthy brain (Mandal et al., 2024).

If disease is thought of as a separate system, it is important to understand how it interacts with the brain. There are various ways in which the interaction between networked systems can be understood. Disease can for instance be thought of as a process unfolding on network structure, which can be static or have its own timescales. Moreover, it is straightforward to conceive of the brain-disease interaction as competitive, though other forms of interaction are in principle possible. Insofar as both brain and disease may possess a network structure, the interaction can ultimately be understood as network competition. However, the interaction need not always be of a competitive nature and the two systems may for instance sometimes better be thought of as coevolving.

*5.1. Disease as a process on a network*

Up until now, we have considered how disease may be associated with changes in network structure. However, various pathologies, e.g. neurodegenerative diseases, may induce not only dynamics *of* networks but also dynamics *on* them, some diseases potentially presenting both kinds of effects (Raj and Powell, 2018). Disease can then be thought of as a process (or modification of a process) unfolding on the structure, the time scales of which are typically different from those of its constituent parts' dynamics. The interplay between dynamical processes and the structure on which they unfold induces novel scales. For instance, dynamical processes induce specific latent geometries, which influence how activity or information may spread across the system (Brockmann and Helbing, 2013; De Domenico, 2017; Barzon et al., 2024).

*5.1.1. The role of network structure in disease spreading*

Various brain pathologies present a distinctive disease progression pattern, with stereotypical neuronal events and corresponding characteristic temporal and spatial scales. For instance, neurodegenerative disorders display highly stereotyped patterns of disease progression, from the entorhinal cortex and hippocampus to the temporal, parietal, and eventually frontal regions (Buckner et al., 2005), whose progression follows fibre pathways rather than proximity (Seeley et al., 2009), characterised by atrophy, tau and amyloid-b pathology, and metabolic load.

Despite fibre tract damage, and evidence that amyloid progression stages are associated with network structure changes between default-mode network brain areas (Pereira et al., 2018), at least some macroscopic aspects of the anatomical network topology in AD is preserved (Powell et al., 2018). Neuropathological evidence suggests that dementias, whose onset is associated with brain atrophy as well as misfolded beta amyloid and tau protein accumulation in the grey matter, may be characterised by spreading along spared white matter tracts through prion-like trans-synaptic transmission of misfolded tau and beta amyloid (Brundin et al., 2010; Raj et al., 2012; Clavaguera et al., 2015). To understand how this process interacts with the network structure it unfolds on, it is necessary to characterise the process itself, at both neuronal and dynamical level, e.g. via passive diffusion, advection, active axonal transport, or other distance-independent processes (Braak and Braak, 1991; Braak, 2003; Iturria-Medina et al., 2014, 2017; Raj and Powell, 2018; Peraza et al., 2019; Del Tredici and Braak, 2020; Pandya and Patani 2021; Millán et al., 2022; Rapisardi et al., 2022). In the former case, misfolded and aggregated proteins may spread through distance-dependent spatial gradient-driven processes (Warren et al., 2013), and the underlying protein transmission process can be modelled as a geometry-driven heat equation (Raj et al., 2012). In the latter case, overall fibre density, corresponding to connection strength at coarse-grained scales, should be more important than distance–dependent spatial diffusion. The atrophy patterns of protein transmission in various dementias were predicted by the Laplacian eigenmodes associated with the dynamics predicted by connectivity-driven spreading models (Stumpf et al., 2000; Matthäus, 2006), consistent with the notion that regional vulnerability to AD is predicted more by network connectivity than by the expression of AD-related genes and the molecular factors which they promote.

A similar use of the network has been proposed to apply to disease progression in schizophrenia. While schizophrenia is characterised by progressive waves of tissue loss (van Haren et al., 2008), with atrophy and cortical thinning (DeLisi et al., 1997; Kubota et al., 2013), dynamic interregional connectivity changes affect the way activity spreads on the anatomical network (Abdelnour et al., 2014) eventually disrupting temporal communication between higher-order brain regions (Hunt et al., 2017). On the other hand, while atrophy dynamics in temporal lobe epilepsy (Bonilha et al., 2003; Bernhardt et al., 2009; Mueller et al., 2009) could in principle result from epileptogenic hyperactivity propagation (Sutula et al., 2003; Riederer et al., 2008) resulting from excitotoxicity (Mehta et al., 2013), it is likelier the result of network dynamics associated with progressive deafferentation, followed by progressive neuronal loss in connected remote regions (Abdelnour et al., 2014, 2015).

*5.1.2. The role of network structure in pathological processes at fast scales*

In some pathologies, e.g. epilepsy or multiple sclerosis, in addition to the disease progression timescale, processes may also unfold at the shorter timescales of disease episodes. One important question is whether seizure-related activity follows the same path as physiological activity (Zaveri et al., 2020). While seizure initiation and propagation have been shown in association with novel local pathways (Bragin et al., 2000; Scharfman et al., 2003), and large-scale anatomical network reorganisation in pharmacoresistant epilepsy patients are common though not ubiquitous (Tellez-Zenteno et al., 2010), seizure propagation has also been shown to use the same pathways of healthy activity in rodents (Rossi et al., 2017). Thus, seizure spreading may somehow depend on the existing network topology and may not be related to atrophy (Liu et al., 2003).

At the single episode scale, spreading is not the only way in which the network can be used by a neural process. For instance, insofar as epilepsy can be understood not only at the spread of hyperexcitability but also as a synchronisation process on the anatomic network, the network structure may play a role in the way the system synchronises, ultimately giving rise to seizure phenomenology (Dyhrfjeld-Johnsen et al., 2007; Chavez et al., 2010; Lehnertz et al., 2023).

*5.1.3. Universal mechanisms in brain disease dynamics*

Brain diseases vastly differ from each other not only in aetiology and dynamical and functional phenomenology but also in their spatial characteristics. In spite of such profound differences there may also exist some universal dynamical traits or mechanisms shared by at least groups of pathologies. For instance, brain injury may be characterised by the spreading along the topological structure of its anatomical connectivity of sleep-like cortical dynamics, resulting from disease-induced local changes in ascending input and lateral excitation and the consequent excitation/inhibition imbalance (Massimini et al., 2024). Likewise, while neurodegenerative diseases feature distinct patterns of selective vulnerability of some classes of neurons, characterised by neuronal loss and protein accumulation, a phenomenon known as *pathoklisis* (Eser et al., 2018), dementias may share a common progression mechanism irrespective of aetiology, region-specific neuropathy (Braak et al., 2000; Pandya and Patani, 2021), or selective vulnerability within dissociated functional networks (Seeley et al., 2009).

*5.2. Brain-disease interactions*

Once brain and disease are thought of as separate networked systems it is natural to assume that they may interact. The interaction between two networked subsystems can have various profound effects. For instance, the interaction of a given subgraph with other nodes in the network affects whether that subgraph corresponds to a fixed-point support (Morrison and Curto, 2019).

What the effects of interactions between networks may be depends on the way they interact. Such a relationship may be understood in terms of interdependence, competition, or even cooperation between network structures (Aguirre et al., 2013; Mišić et al., 2015; Danziger et al., 2019) or coevolution. For instance, interdependence and competition may respectively correspond to positive and negative correlations in the coupling of a system's local connectivity with the other system's local order (Danziger et al., 2019). Whether this simply has dynamical significance or a functional one as well depends on how local order reflects the instantaneous local functionality.

If disease can effectively be considered as a separate networked system interacting with the brain network, one potentially important factor is the way these two networks make contact with each other. This may have two potentially interesting aspects. First, connector nodes may play an important role in the brain-disease interaction (Aguirre et al., 2013, 2014; Buldú et al., 2016). More generally, the interface may itself possess non-trivial network structure, so that disease stages may be quantified in terms of the interface structure and its dynamics.

Another important but still poorly understood question is represented by the quantity these systems may compete for and how this can be expressed in network terms, some form of centrality representing a possible candidate (Aguirre et al., 2013). For example, the activity of the two brain hemispheres has been described as competition process for gaining centrality in the functional network (Martínez et al., 2018a). An analysis of the eigenvector centrality across different frequency bands and conditions showed that both hemispheres generally maintain a functional balance, with no significant centrality dominance of one hemisphere over the other. However, when networks were made sparser, this balance shifted, leading to scenarios where one hemisphere could accumulate disproportionately higher centrality.

## 6. Confronting disease: the role of network structure

One important question is whether subjects' proneness and response to and recovery from disease is related to the system's network properties prior to its inception. Likewise, the structure of or induced by disease may create dynamical traps, which the system cannot easily escape from (Scheffer et al., 2024). The underlying idea is that network structure may not only act as a functional substrate but also as a mechanism of resistance to disease (Fornito et al., 2015). In this sense, the resilience of the anatomical network, in many cases sustained by a high degree of degeneracy, is crucial to maintain the brain within a healthy behaviour (Tononi et al., 1999; Zamora-López et al., 2010). Multiple levels of degeneracy have been reported in the human brain, from microcircuits to large-scale connectomics (Marder and Goaillard, 2006) and has been suggested as one of the key elements behind cognitive reserve and, consequently, a healthy aging (Barulli & Stern, 2013).

The possible role of network structure in resilience can be understood by considering a networked systems described by a global order parameter, e.g. its dynamics $\Gamma$ or information capacity $\mathcal{I}$, depends in a complex way on a network structure $C$, e.g. on the adjacency matrix $\mathcal{A} = A_{ij}$, or on some topological property $\mathcal{T}$ defined on such a matrix, and which lives in a space $\mathcal{S}$, e.g. a manifold $\mathcal{M}$ with some metric $g$. Perhaps the most intuitive aspect one may investigate is the dynamic manifold's smoothness with respect to the connectivity matrix ($\partial \mathcal{M}/\partial C$). A further step may consist in finding how the manifold varies as some particular topological property changes ($\partial \mathcal{M}/\partial \mathcal{T}$). When appraising resilience, though, what is usually addressed is how topology in the appropriate space itself varies with changes in connectivity ($\partial \mathcal{T}/\partial C$). The ultimate and more complex goal is understanding variations induced in the corresponding functional manifold ($\partial \mathcal{F}(\mathcal{M})/\partial C$), and how these may occur as a result in changes in the system's structure.

Addressing resilience from a network perspective involves various important questions. Is there any structural characteristic that renders the system more vulnerable prior to or as a consequence of disease? How much perturbation can the system withstand to maintain structure? Can function be sustained without qualitatively changing structure? How can the system change in order to maintain function without changing structure? Can a system maintain function by changing structure? How much should a system change to maintain function? To what extent does the system's ability to revert to its prior configuration depend on the network structure?

*6.1. Structural, dynamical and functional resilience*

Resilience may in principle be referred to structural, dynamical, and functional aspects of a system (Lesne, 2008; Majhi et al., 2024). Brain *structural robustness* is defined as the degree to which network properties are resilient to such sequences of such perturbations, typically of the anatomical network (Alstott et al., 2009; Warren et al., 2014; Aerts et al., 2016). *Dynamical robustness* is the ability of a system to conserve the same asymptotic dynamics (Lesne, 2008; Demongeot et al., 2010). In particular, it is important to understand dynamical robustness of networks of coupled oscillators (Daido and Nakanishi, 2004, 2007) in biological systems (Tanaka et al., 2015). Insofar as dynamic oscillatory activity is crucial to healthy brain functioning (Başar, 1999; Fries, 2005) and compromised in various brain pathologies (Schnitzler and Gross, 2007), a meaningful way to define dynamical robustness is as the ability of a network to sustain collective macroscopic oscillations when some of its nodes fail to produce rhythmic dynamics as the system is perturbed locally (Tanaka et al., 2012). Note that structural robustness may refer to both the anatomical and the dynamical structure and more generally to any aspect that can meaningfully be endowed with a network structure. Likewise, dynamical robustness could in principle refer to the anatomical structure at sufficiently long time scales. Given brain activity's complex phase space and the non-trivial structure-dynamics-function map, structural and dynamical robustness may or may not imply *functional robustness* (Kitano, 2007; Schwarze et al., 2024). In particular, function may not be robust with respect to given perturbations which leave structure and/or dynamics unchanged.

*6.2. Gauging the effects of disease: the solid material metaphor*

Disease may lead to various degrees of impairment, ranging from decreased efficiency to complete breakdown of a given function. To understand disease and the associated damage in terms of some intrinsic properties of the affected system it is useful to think of the structure, whether defined in real or in phase space, as a solid material. In material science, resilience is related to the amount of energy a solid material can absorb and still recover its original state. This corresponds to its ability to remain in the elastic region of the *stress-strain curve*, obtained by gradually applying load and measuring the deformation on the tested material. This curve reveals various key properties of a material. Two important landmarks in the curve are represented by the *yield point* and the *ultimate tensile strength*. The former identifies the limit between elastic and plastic behaviour, below which the material deforms elastically returning to its original shape when the applied stress is removed, while it deforms irreversibly and plastically above it. The latter identifies the maximum stress that a material can withstand before breaking as it is being stretched or pulled, which lies close to the yield point for brittle materials and further away for ductile ones.

Altogether, the way the system withstands the impact of damage can be thought of in terms of *toughness*. Toughness quantifies the amount of energy that a material can absorb plastically before fracturing, and comprises both *strength*, i.e. the ability to withstand external forces without breaking or yielding and *ductility*, i.e. the ability to withstand plastic deformation before fracture.

Importantly, some properties (e.g. *stiffness*, a measure of a material's ability to return to its original form after being acted upon by an external force), depend on the system's shape and boundaries, while other properties (e.g. the *elastic modulus*, a measure of resistance to elastic deformation) are intrinsic (intensive) properties of a given material.

### 6.3. Resilience in the elastic regime: disease as brain perturbation

One general way to approach resilience and related concepts is to think of disease as a perturbation to the system and to evaluate how the system modulates the response to such a perturbation. Disease can then be thought of as a perturbation acting upon the network structure of a multi-body dynamical system and shoving it towards dysfunctional regimes (Sanz Perl et al., 2023), and vulnerability and resilience are expressed in terms of action upon a network.

Key questions a perturbative model of resilience and pathology addresses are: 1) what aspect is showing robustness and against what kind of perturbation? 2) What aspect of the network structure does disease act upon? 3) Through what mechanism does disease act on the network? 4) How does the network respond to perturbations? 5) How is damage evaluated?

#### 6.3.1. Perturbing brain network structure

Diseases are often represented as factors, e.g. errors or attacks, acting on network node or links, e.g. by deleting, weakening or perturbing them. These factors may come in two main modalities: random or targeted (Albert et al., 2000). In the simplest case, one considers node or link failure, corresponding to their respective deletion. While the effect of network structure can be separated from that of strength (Allesina and Tang, 2012), the impact of link failure on global network function is not related in a simple way to the latter. For instance, the breakdown of strong links may not be critical to network operation, whereas moderately weighted links may induce large-scale performance impairment (Witthaut et al., 2016). Moreover, weak links may have a stabilising effect (Csermely, 2004). Overall, which links are critical is determined by the interplay between redundant capacity, a topological variable, and load distribution (Witthaut et al., 2016).

The consequences of structure perturbation depend not only on the perturbation but also on the structure itself. For instance, they show strong topology-dependence. Topology-dependent sensitivity to random and targeted attacks in isolated networks has long been highlighted (Albert et al., 2000). For instance, it has recently been shown that various dynamical processes unfolding on hierarchical directed networks can be disrupted by targeted attacks to a small fraction of links flowing from a higher to a lower hierarchical level (Rodgers et al., 2023). Moreover, different types of interdependent networks appear to be significantly more vulnerable than isolated networks, featuring first-order (Buldyrev et al., 2010; Parshani et al., 2010; Gao et al., 2011, 2012; Baxter et al., 2012; Radicchi and Arenas, 2013; Liu et al., 2016; Gross et al., 2023a) or hybrid (Liu et al., 2016) phase transitions at link lengths where the mutual giant component still emerges continuously (Gross et al., 2021) in contrast with the second-order continuous phase transition found in isolated networks. Moreover, given the interconnected nature of brain anatomy and dynamics, these results would *prima facie* predict high brain instability. However, the stability of a system of networks results from the interplay between the internal network structure and its connectivity to other networks. In particular, the system of networks is stable and robust to failure if it is interconnected through network hubs, and the connections between networks are moderately assortative (Reis et al., 2014). Insofar as the effects of perturbations are network structure-specific, robustness assessment requires a good characterisation of connectivity at the appropriate scale. In addition to topology at various scales, network geometry can also determine the rate at which a network returns to its original state after a perturbation. In particular, the static anatomical network structure's Ollivier-Ricci curvature can in principle be thought of as a proxy of structural network robustness, as a large curvature is associated with high relaxation rate to the equilibrium distribution (Sandhu et al., 2015).

The standard resilience framework is predicated upon the principle that disease acts through disconnection. Damage is thought of as network dismantling, i.e. how structure fragments into disconnected subsystems as a result of node or connection damage (Braunstein et al., 2016; Ren et al., 2019; Liu et al., 2022), and the functional effects of disease would be achieved by acting on some property of the network to achieve disconnection. For instance, in a directed graph, disease may remove the *feedback vertex set* i.e. a subset of nodes that make the graph acyclic (Zhang et al., 2021). More generally, disease may act on the network's *determining nodes*, i.e. the subset of nodes whose dynamics is sufficient to determine the entire system's dynamics (Mochizuki et al., 2013).

Node or link loss is often modelled as a *percolation* process. For unweighted networks, such a process may involve units failing uniformly at random with some given probability. Alternative failure mechanisms, corresponding to different percolation processes may also exist. For instance, $k$-core percolation involves the iterative removal of nodes with a degree smaller than $k$ (Lahav et al., 2016; Xue et al., 2024). Node and link failure mechanisms have important dynamical consequences. For instance, in $k$-core percolation on spatially embedded networks, link length can induce a metastable phase not present in standard percolation, wherein localised failure spontaneously propagates until the system collapses (Xue et al., 2024). Percolation has been used to probe neuronal resilience in the early stages of brain development. But, while percolation is an elegant way to investigate network properties can it also be understood as a disease's *modus operandi*?

An important issue concerns the deletion process and the extent to which connectedness (particularly anatomical) entails functionality. While intuitive in contexts such as transport or communication, this is not necessarily the case for all brain functions. Indeed, anatomical fragmentation is not a necessary condition for brain function impairment and functional breakdown may precede structural fragmentation. For instance, generalised or focal seizures may arise in the absence of localised brain damage (Terry et al., 2012). Moreover, removal of nodes or links central to network anatomical connectivity might have insignificant effects. More generally, pathology is indeed more generally defined by *dysconnection* rather than *disconnection*, as link strength and density may also be enhanced (Friston et al., 2016). Functional impairment may also result from failure to activate an otherwise intact anatomical structure. This may stem from noise on nodes but also from noise on links (Cavagna et al., 2024). Brain damage may also act by impairing cross-scale activity flow, which can be studied by examining the response to damage of dynamical processes (Ghavasieh et al., 2023).

A further important issue is the representation of functional damage in terms of some aspect of network structure. The impact of a structural perturbation is typically quantified in terms of the largest connected component, i.e. the subnetwork that remains connected following after the random removal of a fraction of its nodes, as a function of the pruning probability (Rapisardi et al., 2022). Network robustness may for instance be characterised in terms of the size the giant component or of the integrated size of the largest connected cluster during the entire attack process (Gao et al., 2015) or of the critical threshold value at which the giant component vanishes and the network becomes disconnected (Liu et al., 2022). Crucially, these criteria can be expressed in terms of structural properties of the network. For instance, for a random network, the critical threshold can be expressed as a function of the degree distribution (Cohen et al., 2000). The size of the largest connected component may for instance undergo phase transitions at different thresholds or even of different types. But what physiological meaning should be ascribed to the largest connected component? It is reasonable to suppose that its meaning should be scale-dependent (i.e. it should only be meaningful at some scales). Moreover, connectedness provides no information on network connectivity's resilience before or after attacks have been performed (Mohseni-Kabir et al., 2021). Alternative criteria may be more sensible depending on the context. For instance, in transport, a more appropriate criterion is for a node to belong to the network's backbone, where information flow, or blood supply occurs. However, it is in general not straightforward to evaluate a given networked system's resilience in terms of the giant component *per se*, and other network properties, e.g. modularity, feedback loops (Csete and Doyle, 2002) or some non-trivial network property possibly non-local in the real space may be more appropriate indicators of functional resilience.

Alternative methods to evaluate stability are sometimes used. For instance, the relation between anatomical and dynamical network structure is often used to predict the amount of damage the brain can withstand due to lesions at a given location. While this may provide some indirect indication it is nonetheless not a clear indication of functional robustness. Moreover, structural perturbations can also be used to probe dynamical rather than structural robustness (Faci-Lázaro et al., 2022).

Rather than in structural terms, resilience against structural perturbations should perhaps instead be evaluated in functional terms, e.g. in terms of information transport or information coding and capacity. For instance, error-correcting codes, which encode information so as to guarantee its faithful transmission even in the presence of various sources of noise and codes maximising the capacity of a network as a whole should differ in underlying network structure they induce, but also in terms of robustness, including the type of perturbation they are designed to cope with, and possibly in terms of the network adaptation following node or link failure that they require (Koetter and Médard, 2003; Koetter et al., 2011). Thus, it is expected that coding strategy and network structure interact to modify the functional consequences of perturbations. More generally, network structure may mediate trade-offs between robustness resource demands and performance. For instance, redundancy enhances robustness against component failure, but for a constant resource level, it may degrade performance. It may also increase vulnerability against other types of perturbation (Kitano, 2007).

*6.3.2. Perturbing brain network dynamics*

Instead of acting upon the system through node or link deletion, disease may challenge the system via a proper matrix perturbation through infinitesimally changes. While the former effectively changes the connectivity matrix's dimensionality and therefore constitutes a matrix change, the latter constitutes a genuine matrix perturbation.

Perhaps the simplest scenario is one in which oscillatory activity at node level gradually transitions into fixed points (Daido and Nakanishi, 2004). If enough nodes undergo such a transition, the system may transition towards a globally non-oscillatory regime, a phenomenon termed *ageing transition* (Daido and Nakanishi, 2004). In a global network of diffusively coupled oscillators, aging transition can be characterised by a universal scaling law of an order parameter controlled by the fraction of transitioning nodes and the coupling strength (Mahji et al., 2024). Ageing transition has been studied in various models with different coupling functions and network structures (Tanaka et al., 2014; Pazó and Montbrió, 2006; Thakur et al., 2014). For instance, it has been shown that scale-free networks are dynamically robust with respect to random inactivation but vulnerable to targeted inactivation of low-degree oscillators. Remarkably, this stands in contrast with the role hubs play in scale-free networks' structural robustness. Furthermore, the presence of time delays in the coupling decreases dynamical robustness (Thakur et al., 2014), whereas dense network motifs enhance dynamical stability (Gross et al., 2023b). A further important question is what makes synchronous states more or less fragile against external perturbations (Tyloo et al., 2018).

An important point is to do with the evaluation of resilience. Often, for instance, when intended as a measure of stability, resilience is evaluated via local linear measures and against infinitesimal perturbations. Brain dynamical resilience can for instance be defined as the minimum distance between all accessible dynamical regimes (Rings et al., 2019a). The larger this distance the higher brain's ability to withstand perturbation and reorganise undergoing dynamical changes while retaining the same functionality. Dynamical resilience was found to increase in the hours preceding seizures (Rings et al., 2019a). However, to evaluate stability against generic perturbations, non-local and nonlinear measures such as basins of attraction's volume, a measure of the likelihood of return to a given state after random not necessarily infinitesimal perturbation, may be preferable (Menck et al., 2013).

*6.4. Network structure and resilience mechanisms in the elastic range*

To understand the role played by brain network structure in disease prevention or persistence it is essential to characterise the mechanisms through which the system acts to enforce robustness. Robustness may for instance result from redundancy or degeneracy (Rule et al., 2019; Jensen et al., 2022) and *sloppiness*, i.e. the dynamics may be sensitive to but a few combinations of parameters and insensitive to many other combinations (Daniels et al., 2008; Transtrum et al., 2015; Panas et al., 2015). Functional redundancy and degeneracy arise at many levels (Edelmann and Gally, 2001), although how they are organised and interact across scales is still poorly understood (Machta et al., 2013). Robustness has also been related to feedback and control mechanisms (Barkai and Leibler, 1997; Doyle and Csete, 2005). The main question here is what role network structure plays in the implementation of such mechanisms in the healthy brain and in disease: does structure ensure robustness of certain properties against a range of perturbations? Conversely, is network structure an emergent property of trade-offs of systems favouring robustness?

Network structure may contribute to robustness in various function- and perturbation-specific ways (Schwarze et al., 2024), in both real and phase space. For example, higher-order interactions in real space may stabilise dynamics (Grilli et al., 2017). Network structure may also somehow constitute a sloppy parameter, although exactly how, at what scale and what properties should be determined. A further possibility is that network structure could provide topological protection (Murugan et al., 2016), characterised by insensitive of fundamental properties of brain structure, e.g. modularity, dynamics or function to almost to smooth changes in parameters, with change requiring a phase transition (Hasan and Kane, 2010), due to integer-valued quantities called *topological indices* (Thouless et al., 1982; Wen, 1995). A classical example is the *genus* of a surface, which counts the number of holes in the surface (Ghrist, 2014). In phase space, the relational structure could constitute an obstruction to some functionally important property, e.g. convexity in neural codes (Ghrist, 2014; Giusti and Itskov, 2014; Curto et al., 2017). For instance, physiological properties of neuronal populations may be associated with code convexity, while codes with local obstructions are necessarily non-convex (Giusti and Itskov, 2014; Curto et al., 2017; Lienkaemper et al., 2017).

*6.4.1. Neutral structure*

In many biological systems, e.g. in gene regulatory networks (Bergman and Siegal, 2003; Ciliberti et al., 2007), resilience results from the presence of *neutral networks* (van Nimwegen et al., 1999; Schuster and Fontana, 1999; Wagner, 2008a) and the associated *neutral space*, i.e. the corresponding subspace in the phase space. In a genetic context, a neutral network is a set of equifunctional genes related by a point mutation, whose nodes represent gene sequences and whose links are mutations connecting them (van Nimwegen et al., 1999). Thus, the structure in the functional space is induced by the genotype-to-phenotype map and by the mechanisms controlling mutations. Similarly, in the context of RNA secondary structure, RNA sequences are mapped onto the genotype and the secondary structures onto which they fold play the role of phenotypes and what is considered is the phenotypic structure induced by the sequence-to-structure map (Schuster and Fontana, 1999; Wagner, 2008a).

In a neural context, a neutral network can be defined at various levels. For instance, it is natural to think of the physiological space as the genotype space, while the phenotype space may be identified with the space of network structures, e.g. of topologies, induced by the renormalisation of the underlying physiology. This space is both scale- and renormalisation-dependent. Finally, the functional space is identified with the functional or fitness space. Overall, one considers genotype-to-fitness maps (Stadler et al., 2001; Manrubia et al., 2021), which can be decomposed into a physiology-to-structure and structure-to-function map, highlighting the role played by the intermediate phenotype space. A neutral structure can in principle be defined at both genotype and phenotype levels. In the latter case, the neutral network is a structure whose nodes are network ensembles resulting in functionally equivalent phenotypes, and whose links are induced by the associated accessibility structure. The corresponding neutral space comprises regions in the space of network structures (or of parameters) that give rise to equivalent functional behaviour.

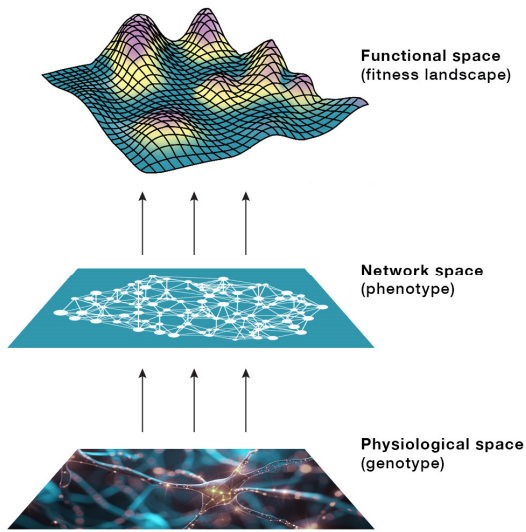

**Fig. 3.** Functional brain activity can be thought of as the fitness landscape corresponding to a given network structure phenotype. In turn, the phenotype space of network structures emerges from the renormalization of underlying space of physiological activity. Neutral structure can be defined at both genotype and phenotype level. The functional space's morphology depends on the properties of both the genotype-to-phenotype and phenotype-to-fitness maps.

Phenotypes' accessibility is determined by the interplay between the mechanisms generating variation and the structure of the genotype-to-phenotype map (see Figure 3). If functional phenotypes are organised according to accessibility of given network structure ensembles, the resulting functional space may have a non-trivial structure. Note that, at both the genotype and the phenotype level, the neutral network is not defined in real space but rather in phase space, although each node may itself represent a topological pattern in real space renormalized up to a given scale and phenotypic neighbourhoods, which are induced by the genotype-to-phenotype map, may considerably differ from neighbourhoods in the real space. Robustness implies that the different genotypes of a given phenotype are close in the genotype space. Moreover, phenotypic transitions can be thought of as competitions between networks (Manrubia et al., 2021).

The presence of neutral networks may have important functions, e.g. allowing communication in the presence of noise as in error-correcting codes (Hopfield, 1974; Sreenivasan and Fiete, 2011), and exploring the configuration space without dysfunctional phenotype changes.

### 6.4.2. Extension of the elastic range: cognitive reserve and symptom onset

Often, a temporal gap separates disease onset from the appearance of observable symptoms, and a further time delay exists between symptom onset and complete functional breakdown. For instance, AD symptoms can be delayed in subjects with a high educational level in spite of neuropathological changes. It has been proposed that this phenomenon may be explained in terms of *cognitive reserve* (Mortimer, 1988; Le Carret et al., 2003; Buckner, 2004; Stern et al., 2019; Wilson et al., 2019). Following the material metaphor, how cognitive reserve shields the system with respect to symptomatic functional disruption could be understood in terms of extension of the elastic range width. One way to understand whether and how functional and behavioural changes are related to network structure may involve characterising the resilience of brain networks particularly of the hierarchical structure in ageing (Stanford et al., 2022), e.g. through $k$-shell decomposition [2] (Dorogovtsev et al., 2006). Core connectivity and resilience was found to decline with age (Stanford et al., 2022). Furthermore, differences in the organization of functional networks among individuals with varying levels of cognitive reserve have been linked to the dynamical properties of specific brain regions. Notably, regions exhibiting higher connectivity strength tend to show increased entropy and reduced complexity, reflecting more stochastic and less organized neural dynamics within network hubs (Martínez et al., 2018b). On the other hand, cognitive reserve has been shown to be associated with increased redundancy in brain networks as quantified by the number of simple (non-circular) paths between a pair of nodes up to a specified length (Stanford et al., 2024). Fundamental questions are to do with the specificity but also with irreducibility of such a result. For instance, what is attributed to $k$-shell topology may in fact reduce to lower-level properties such as degree distribution (Baxter et al., 2015). Finally, it is also unclear through what mechanism and against what class of perturbations particular network structure enforces the resilience of neural systems associated with cognitive reserve.

### 6.5. The role of network structure in brain reorganisation

As disease pushes the brain away from its neurophysiological structure, various dynamical scenarios are possible. First, the system may be able to reverse disease and self-heal. This could drive the system to the original attractive state or to some equifunctional albeit different one from the one where the system was prior to disease. Quasi-neutral regimes may not be neighbours in the genotype and in the phenotype space, i.e. in the physiological space and in the way this renormalizes into a particular network structure ensemble. They may for instance result for the recruitment of new degrees of freedom at the neurophysiological level, resulting in topologically different but equifunctional structure. On the opposite side of the spectrum, disease may be irreversible, though not lethal, and become chronic. This does not necessarily entail that it dwells in the dynamical regime directly induced by the disease. For instance, following adverse events such as strokes, the brain may head towards parts of the phase space which, though not equifunctional with respect to those corresponding to healthy functioning, are nonetheless functionally improved with respect to those of the acute phase. In doing so, the neuronal populations explore the neurophysiological configuration space. In such exploration the system can evolve and reorganise at scales often much larger than those of the perturbation, either slowly adapting through evolution or directly through phenotypic plasticity (Botero et al., 2015). In this regime, resilience may involve profound changes in some aspects of brain structure and dynamics. The exploration process can be characterised in various ways. For instance, once the brain has experienced the adverse functional effects of disease it can be thought to undergo learning to perform a new function in search of a fitness maximum (Stern et al., 2024). In a networked system, learning is typically associated with changes in the network structure (Stern and Murugan, 2023), and the corresponding neural activity is encoded in low-dimensional manifolds (Gao and Ganguli, 2015).

---

[2] The $k$-core of a graph $G$ is the maximal subgraph of $G$ having minimum degree at least $k$. The $k$-shell of a graph $G$ is the set of all nodes belonging to the $k$-core of $G$ but not to the $(k+1)$-core. The network is decomposed into node shells of increasing interconnectivity robustness, until a core set of nodes within a network is reached (Dorogovtsev et al., 2006). Attacks are performed on these networks, and the resilience of nodal shell and core are measured by extracting the k-shell of the resulting network (Mohseni-Kabir et al., 2021; Shang, 2019; Wang et al., 2022).

Several important questions arise: what is the dynamics of such wandering process? By what is it driven? Where does it lead to, i.e. what is the nature of the states or regimes it leads to? In all cases, what is the role of network structure? Is recovery associated with the attainment of a particular network structure? Conversely, is a particular network structure associated with the system's ability to evolve towards a state with improved function?

*6.5.1. The role of network structure in brain evolvability after injury*

Insofar as disease can effectively be thought to exert effects similar to those of a mutation more or less abruptly shoving the system towards non-neutral (possibly non-neighbouring) parts of the phase space (Kaneko, 2007; Ciliberto et al., 2007; Greenbury et al., 2016), one important question relates to the system's ability to evolve, and improve functionality in particular through rewiring and changes in dynamics (Altenberg, 1995; Payne and Wagner, 2019). In this sense, the brain should be considered as an evolutionary capacitor buffering genotypic variations (Bergman and Siegal, 2003) and disease may be thought of as a loss not only of fitness and resilience, but also of evolvability. But what role does network structure play in brain evolvability? In particular, does disease affect evolvability by acting on network structure? Does disease reveal physiologically *dormant phenotypes*?

An essential ingredient for evolution is the heterogeneity in traits affecting function. In genetics, evolvability is quantified in terms of the variety of phenotypes lying within a given mutation distance of a genotype or phenotype (Wagner, 2008b). If disease drags the system to some suboptimal genotype, an important aspect is therefore *phenotypic variability*, i.e. the extent of phenotypic variation accessible to the genotype. A given genotype can either be robust, when surrounded by neighbours of the same phenotype, or evolvable, when its neighbourhood comprises a variety of other phenotypes, phenotypic robustness may actually facilitate evolvability (Wagner, 2008b, 2011, 2012; Ahnert, 2017), due to the presence of neutral networks and *genotype neighbourhoods*, i.e. the set of genotypes accessible from a genotype in a given number of mutations, a measure of genotype's phenotypical variability (Wagner, 2008a). Genotypes within a neutral network can thereby explore different phenotypes. Translated in brain network terms, this means that neurophysiological activity should be able to renormalize into separable network ensembles.

A further fundamental question is whether there are network properties facilitating evolvability. In both real and phase space, an important property facilitating evolvability is represented by modular organisation (Simon, 1962). Interactions between different components provide a phenotypic space where mutations can produce variation without inducing adverse consequences. In phase space, phenotypic robustness and evolvability can be characterised in terms of network properties (Aguirre et al., 2018). For instance, the former can be defined through topological properties of the phenotype or of the functional space, the latter as the emergence of connected components of the neurophysiological space. In turn, this definition suggests that *robustness of evolvability*, may itself be a robust property, which can be quantified as the sensitivity to network perturbations of the topological measure quantifying evolvability (Ibañez-Marcelo et al., 2014).

Finally, it is worth noting that if disease is a phenotype with its own structure, it should also be characterised by its own resilience and evolvability. For instance, cancer can be seen as a system robust with respect to its proliferative potential, which is robust against the action of the immune system and therapies (Kitano, 2004a,b), but also highly evolvable (Tian et al., 2011), so that effective control strategies must address both robustness and evolvability (Tian et al., 2011).

*6.5.2. Beyond the elastic range: learning, homeostasis, and adaptive resilience*

Up until now, we have considered disease through an approach wherein a structure responds to some particular perturbation in an essentially passive way. In network terms, the proximal effects of such a perturbation are typically translated into local node or link damage, ranging from perturbation to break-down, which may induce cascades of secondary effects. In this framework, disease onset and functional breakdown are respectively thought of as the system's yield point and optimal tensile strength. However, this approach does not take into account living systems' fundamental tendency to maintain a given function rather than a given state or regime e.g. a given network structure (Kitano, 2007). Moreover, we have considered resilience with respect to perturbations from elements acting on the system but which are not modified by it. The picture is bound to be different when brain and disease can effectively be thought of as interacting systems. In the same way that disease may directly affect the underlying network structure associated with healthy brain function and achieved during the course of development and learning prior to disease onset, the dynamics following disease may also affect (and be affected by) the network structure induced by the disease. This is particularly conspicuous in chronic disease.

The ability to absorb shocks and reorganise so as to retain the same function requires plastic mechanisms in a wide range of scales, from the fast scales of sensory stimuli, to developmental and evolutionary ones, enabling the system to change beyond the elastic range. Brain function is subject to a trade-off between stability and plasticity (James, 1890). On the one hand, it must produce stable representations to prevent previous knowledge forgetting. Specific network properties in the relevant space may characterise not only resting brain activity over timescales of minutes to hours, but also processes such as memory storage, where representation stability is required in the face of connectivity fluctuations (Susman et al., 2019; Rule et al., 2019). On the other hand, neural systems from cell to population levels must be able to match incoming stimuli, to integrate new knowledge and to learn from past experience in order to improve fitness but also to withstand phasic stress of various types (Mattson and Magnus, 2006; Rutkowski and Hegde, 2010; Roth and Balch, 2011) and to adapt to chronic stress (Saxena and Caroni, 2011). Correspondingly, network structure may play seemingly opposite roles under different circumstances.

Learning and adaptation are subserved by various classes of physiological mechanisms acting at various spatial and temporal scales. For instance, the structure can evolve as the result of mechanisms enabling learning of various forms (Clark and Abbott, 2024). However, an autocatalytic mechanism such as Hebbian learning would lead to runaway excitation and complete network synchronisation (Markram et al., 1997; Ocker et al., 2015; Zenke et al., 2017). A class of plasticity mechanisms capable of achieving dynamic stability in recurrent networks in the presence of Hebbian learning is represented by *homeostatic plasticity*, through which neurons control their own excitability, regulating spike rates or stabilising network dynamics at various time scales (Turrigiano et al., 1998; Turrigiano, 2011). Importantly, in the neocortex, local networks' correlation structure (Wu et al., 2020) and network criticality (Shew et al., 2015; Ma et al., 2019) are under homeostatic regulation (Wen and Turrigiano, 2024; Zeraati et al., 2020; Landmann et al., 2021; Menesse et al., 2022). Plasticity may dynamically target criticality by adjusting excitatory-inhibitory connectivity to address excitation/inhibition imbalances in neuronal populations (Sukenik et al., 2021).

Homeostatic plasticity mechanisms are slow negative feedback loops (Ma et al., 2009; Zierenberg et al., 2018; Golubitsky and Wang, 2020) implemented by several neurophysiological mechanisms. For instance, synaptic scaling (Markram and Tsodyks, 1996; Turrigiano et al., 1998; Fong et al., 2015; De Vivo et al., 2017) allows conserving the overall neural activity in the presence of exogenous perturbations or damage, by adjusting synaptic strength (Murphy and Corbett, 2009), membrane excitability (Davis, 2006; Pozo and Goda, 2010), or neuron-glial interactions (de Pittà et al., 2016).

Through appropriate combinations of plasticity mechanisms, the brain can optimise wiring, by creating or pruning connections, as well as weight distribution (Estévez-Priego et al., 2020; Teller et al., 2020; Schwarze et al., 2024), reshaping the network structure at all scales and generating complex dynamical regimes (de Arcangelis et al., 2006; Levina et al., 2007, 2009; Meisel and Gross, 2009; Rubinov et al., 2011). For instance, interactions between Hebbian learning and single-cell homeostasis allow neurons exploiting redundancy in order to reliably read out an evolving population code in the presence of gradual tuning drift (Rule and O'Leary, 2022). Notably, at least in principle, different plasticity rules can give rise to indistinguishable network dynamics (Ramesh et al., 2023).

Plasticity-induced rewiring can conceivably achieve both improved functional states and higher resilience through various mechanisms. For instance, it can effectively enforce *allostery*, a mechanism whereby an event at one site affects the activity at a distinct site, enabling the regulation of the corresponding function. Furthermore, self-regulatory mechanisms such as neural scaling triggered by damage could adjust link strength through *filtration processes*, whereby all links with a weight strictly below a given threshold are removed (Rapisardi et al., 2022), mitigating through an active process the damage associated with link removal. Notably, these processes may involve parts of the system not confined to the one directly undergoing stress, and cells may proactively summon stress pathways to cope with higher demands. For instance, when cells initiate processes increasing pressure on homeostatic systems they also activate stress pathways to meet the associated physiological needs (Rutkowski and Hegde, 2010).

### 6.5.3. Significance of network structure in functional reorganisation after brain injury

Plasticity mechanisms can change real space network structure in a wide range of scales, but several fundamental issues related to plasticity following disease are still poorly understood. First, while there is no consensus on the potential and limits of functional reorganisation after injury (Makin and Krakauer, 2023), it is unclear whether the potential for recovery associated with some specific set of network properties, particularly in real space. Likewise, no consensus exists as to whether functional reorganisation is a maladaptive response to injury (Penner and Aktas, 2017; Rocca and Filippi, 2017). Second, damage is in general thought to exert a disruptive effect on structure, but it could also in principle directly or indirectly promote functionally relevant non-trivial surrogate structure. For instance, in optimised transport networks damage can induce the formation of loops (Katifori et al., 2010; Kaiser et al., 2020). On the other hand, the structure associated with functional reorganisation need not be functional, particularly past tipping points.

But can disease-induced functional and dysfunctional structure be discriminated in terms of network structure?

### 6.5.4. Structural consequences of aberrant plasticity

Disease can act in two ways: either by overcoming the system's ability to adapt plastically or by interfering with plastic mechanisms themselves. Plasticity is a necessary condition for brain function and both excessive plasticity or dysfunctional homeostasis render the brain unable to adjust to changing demands and excessively vulnerable to environmental perturbations (Pascual-Leone, et al., 2011). Various pathologies, including developmental disorders or neurodegenerative diseases may be related to altered plasticity (Hallett, 2005; Oberman et al., 2014). The disorder phenotype is specified by the particular combination of environmental factors, genetic abnormalities and their expression timing, affected cortical network and synapse type and the direction in which it is affected (Oberman and Pascual-Leone, 2013). For instance, idiopathic ASD can be thought of as a plasticity disorder characterised by altered local cortical excitation/inhibition balance and brain network functional connectivity, resulting from genetic and environmental factors leading to aberrant excitatory plasticity (Oberman et al., 2014). At the cell level, ASD may stem from increased axonal density and complexity and a reduction in synapse pruning and elimination resulting from GABA synthesis blockade (Wu et al., 2012). On the other hand, neurodegenerative diseases may be initiated by chronic, possibly interacting, perturbations acting upon homeostatic mechanisms at various scales. For instance, AD's early stage has been suggested to arise from an imbalance between firing homeostasis and synaptic plasticity (Styr and Slutsky, 2018). Dysfunction could become symptomatic when disease-related stress exceeds the system's ability to withstand it. Disease may be aggravated by systemic feedback, e.g., through inflammation and immune responses, which can ultimately lead to neuronal degeneration and death, and disease spreading. Disease may also affect the hierarchical unfolding of plasticity-related neurobiological events during neural development. For instance, various developmental psychopathologies are characterised by protracted plasticity in association cortices (Sydnor et al., 2021).

But what is the role of network structure in aberrant plasticity? The main question is whether functional impairment in hyperplastic pathologies can be traced back to changes in the associated network structure. For instance, plasticity and adaptiveness may underlie increased pre-seizure dynamical resistance and reduced medication effectiveness (Rings et al., 2019a; Zaveri et al., 2020; Lehnertz et al., 2023).

### 6.5.5. Beyond the plastic range: disease as structure failure and its structural harbingers

Brain disease often leads to complete breakdown of function. In the material metaphor terms, this corresponds to the ultimate tensile stress. An important question is whether this point can be predicted and, if so, whether it is accompanied by specific network properties in real and in phase space.

In strongly disordered non-hierarchical networks, micro-crack accumulation preceding failure is characterised by avalanches with self-affine scaling behaviour. Failure can then be thought of as a critical phenomenon and avalanche size as an order parameter, whose change constitutes a failure precursor. From a dynamical view-point, failure is correspondingly described in terms of either *damage percolation* or *crack nucleation-and-growth*. In contrast, in hierarchical materials, fracture microstructure is generically scale-invariant with self-similar patterns at different scales even very far from the critical load (Moretti and Zeiser, 2019; Moretti et al., 2018, 2019). Failure is announced by two different properties, i.e. increased *eigenvector localisation* and decreased *topological dimension* (Moretti et al., 2018). Importantly, such a structure is associated with diminished crack propagation, containing damage spreading, and with changes in intrinsic material properties such as increased toughness (Moretti et al., 2018).

The material metaphor could conceivably apply to anatomical networks, but can it also be used to make sense of dynamic network resilience? In particular, are brain transition points associated with network biomarkers in brain dynamics? At long timescales, an important problem is whether the transition to fully-fledged neurodegenerative conditions such as AD from its prodromes are associated with and predictable based upon corresponding network features.

At shorter time scales, network-based early-warning signs of imminent bifurcations or sudden deterioration have been reported (Chen et al., 2012; Liu et al., 2015). For instance, although transitions may not be characterised by classical features of criticality (Wilkat et al., 2019), seizures have been shown to be preceded by a decrease of neuronal network resilience which determines the ictogenic potential of interictal synaptic perturbations (Chang et al., 2018). On the other hand, while temporal changes of single nodes and links properties reported were found to be predictive of seizures (Andrzejak et al., 2003), dynamical networks' global properties have been found to hold only limited predictive power (Kuhnert et al., 2010; Takahashi et al., 2012; Geier et al., 2015; Mitsis et al., 2020). At macroscopic scales, seizure precursor generation may result from a rearrangement of the epileptic network's path structure, eventually resulting in the formation of bottlenecks altering activity spreading (Rings et al., 2019b). The earliest indications for such changes may occur hours prior to seizure onset and stem from connectivity changes within and between brain regions distant from the seizure onset zone corresponding changes in the time-varying centrality of the associated nodes. These regions which are part of the large-scale epileptic network generate and sustain physiological brain dynamics during the inter-ictal state also control the dynamics within the seizure onset zone (Lehnertz and Dickten, 2015; Johnson et al., 2023), and may modulate seizure dynamics (Geier et al., 2015; Khambhati et al., 2016). On the other hand, indicators found minutes before a seizure in the spatial proximity to the seizure onset zone appear to be epiphenomena of earlier harbingers of ictogenesis (Rings et al., 2019b).

*6.6. Preventing and curing disease through network-based intervention*

From a clinical viewpoint, it is important not just to understand how the brain copes with pathological perturbations but also to prevent or reverse disease, through therapeutical interventions including pharmacological medication, surgical procedures, brain stimulation, neurofeedback or behavioural therapy (Flanagan et al., 2019; Papo, 2019b). But is it possible to prevent a system from approaching potentially pathological regimes or to shove it away from such a regime once already in that state by acting upon brain network structure? Can resilience be increased (or disease resilience decreased) by acting on network structure?

Available strategies are essentially of two types: lesioning or restorative. The former involve surgery designed to shift the system to a functionally equivalent structure, effectively constituting an *antifragile* strategy, i.e. damaging targeted structure to improve the performance of a global function (Taleb, 2012). The latter comprise stimulation strategies from chronic deep brain stimulation to phasic stimulation and neurofeedback.

A paradigmatic example of network-based intervention strategy is represented by seizure control procedures (Jirsa et al., 2017; Olmi et al., 2019). At the most basic level, seizure relief can sometimes be achieved by limiting seizure propagation, rather than addressing the seizure itself, the underlying principle being that cutting a dysfunctional network into pieces may restore its functionality (Ren et al., 2019). This can be achieved via resective and non-resective surgery, ablation or stimulation strategies (Morrell et al., 1989; Papo et al., 1990; Dogali et al., 1993). On the other hand, while multiple focal ablations within the *epileptogenic zone*, i.e. the portion of cortical tissue necessary and sufficient for seizure initiation (Luders et al., 1993), may completely abolish epileptiform activity, such a strategy may also rebalance the network and hinder seizure propagation when targeting outside the epileptogenic zone (Olmi et al., 2019).

A proper intervention strategy involves defining in addition to a general goal, a neural target as feature, including *what* aspect of brain activity to act upon and *where* in the brain, and an appropriate stimulation schedule (Papo, 2019b). What makes network-based interventions more arduous than standard local interventions? At the most basic level, a purely network-based approach may involve intervention at multiple sites. More substantially, such an approach adds a complex factor to the control of dynamics, implying the identification and specification of a structure, in real or in phase space, as well as an understanding of the dynamical and ultimately the functional role of such a structure in the system's behaviour. Such understanding may exist, at macroscopic scale precisely when the underlying neural system works in a relatively simple way, e.g. as a feedforward loop as in basal ganglia's loops. The extent to which the relevant network structure must be known for the intervention to be effective, e.g. in seizure control, is also a matter of current debate (Zaveri et al., 2015). Network theory suggests various methods to protect a network-of-network from cascading failure, e.g. protecting high-degree nodes against failures and attacks (Schneider et al., 2013), forcing dependency relations between nodes of similar degree (Parshani et al., 2011) or ensuring interconnections of independent parts through nodes with particular degree (Aguirre et al., 2013; Reis et al., 2014). However, at a practical level, the main issue is finding a way to act on the network in accordance. For instance, it was proposed that network connectivity in AD could be preserved by altering neuronal excitability (de Haan et al., 2017). Moreover, intrinsic properties of neuronal circuits such as plasticity or degeneracy can sometimes complicate therapeutic intervention. For instance, changes in several ion channels are sufficient though not necessary conditions for hyperexcitability in primary somatosensory neurons induced by nerve damage in neuropathic pain, so that addressing degeneracy requires an integrative approach to drug discovery (Ratté and Prescott, 2015).

*6.6.1. Control and targeting of brain networks in disease*

It is straightforward to think of medical intervention as a control problem, wherein brain dynamics, pushed away from the neurophysiological regime by disease, must be nudged (in finite time) towards the basin of attraction of a functionally more desirable dynamical regime, to which the system would not (or too slowly) spontaneously evolve. This can be implemented by finding an appropriate control parameter for the dynamics and a feasible path from the initial condition to the target dynamical regime and altering either the system's dynamical equations (Ott et al., 1990; Boccaletti et al., 2000; Lai, 2014) or its initial condition (Cornelius et al., 2013), either stabilising a specific trajectory through the application of small time-dependent perturbations, or steering the dynamics towards a trajectory compatible with the system's natural dynamics but originating from a different initial condition, a procedure called *targeting* (Shinbrot et al., 1990). Moving a spatially-extended networked system into any desired state within a given continuous phase space volume adds further complexity to the control problem associated with the spatial and relational nature of the controlled system (Liu and Barabási, 2016). Both control and targeting are in general conceptually approached through local perturbations of nodes or links.

Control methods could in principle be used for network reprogramming, for example, to shove brain dynamics away from epileptic or epileptogenic regimes (Cornelius et al., 2013), or to promote a particular behaviour of the system (Gutiérrez et al., 2012). More fundamentally, network control theory could in principle be used to probe cognitive resilience (Medaglia et al., 2017; Khona et al., 2022). Within this context, it is straightforward to address the following questions: what dynamical states are accessible in a stable way, starting from a given initial condition? What's the minimal number of nodes or links that need to be perturbed to reach a given desired dynamic regime? How costly is achieving states with given performance levels?

Recently various studies have tried assessing brain controllability (see Bassett et al., 2017 for a review). Often, however, these studies are predicated upon rather unrealistic hypotheses on brain activity: brain resting dynamics is described in terms of a set of differential equations linearised around a dominant fixed point and connectivity dynamics is assumed to be linear and time-invariant (Kim et al., 2018). Control of networked non-linear dynamical systems faces a number of other challenges (Sun and Motter, 2013; Liu and Barabási, 2016; Zañudo et al., 2017; Tu et al., 2018). For instance, control trajectories are in general *non-local* in the phase space, i.e. moving the system from the initial to the target state may require a very long path, irrespective of the distance between these two points in phase space; there is also a non-locality trade-off whereby either the control trajectory is non-local in the phase space or the control inputs are non-local in the network (Sun and Motter, 2013). More generally, ascertaining whether the system is *controllable*, that is, whether it can be driven from any initial condition to any desired state in finite time (Kálmán, 1963) typically requires extensive knowledge of the system's state space, and sometimes of its dynamics (Cornelius et al., 2013), an information usually unavailable for system-level brain activity. Furthermore, given the heterogeneity and multiscale character of brain dynamics, it is highly non-trivial to determine both the overall effects of anatomically local stimulation and the level of coarse-graining which would guarantee good enough control targets. An additional issue is given by control *costs*. In the case of brain stimulation techniques, how much power is needed to achieve control and whether this is compatible with safety are fundamental issues. While network control methods aim at controlling dynamics through a minimal number of nodes or links, too exiguous a number may exact an excessive energetic cost (Yan et al., 2015). Contrary to reports suggesting resting brain activity controllability through a single node representing a given brain region (Gu et al., 2015), a recent study (Tu et al., 2018) showed that even though brain networks might be structurally controllable, the energy required to control it may be disproportionately high.

*6.6.2. Recovering from disease through microscopic interventions*

Often, treatment strategies, particularly those using brain stimulation techniques, are predicated upon the idea that damage can be reversed by reconstructing the healthy regime's structure, e.g. by retrieving damaged nodes or links or, when this is not possible, by inducing a topologically and functionally equivalent one. However, due to the presence of path-dependence, restoring topology, though possibly necessary, is in general not sufficient to return to the functional state. To address this issue, one possible strategy may consist in identifying states in which the system can be acted upon through microscopic interventions, for instance, by merely controlling a few nodes (Sanhedrai et al., 2022; Sanhedrai and Havlin, 2023). This strategy eliminates potential undesired inactive states, keeping the system in its healthy regime, without moving it as in control strategies. Moreover, while control strategies typically act locally, this strategy can act globally in the system's phase space (Sanhedrai et al., 2022).

# 7. Concluding remarks

(1) Over the past few years, it has become standard to think of brain anatomy and dynamics as systems whose non-trivial network structure plays an essential role in healthy brain function. This assumption, predicated upon theoretical studies and experimental results from systems often rather different from neural systems, allows various angles and insights on disease, its consequences and the possible ways it can evolve and be acted upon. Overall, the intrinsic relational structure of brain anatomy and dynamics can be thought to mediate fundamental functional trade-offs between efficiency, functional reliability, and evolvability, which ultimately determine both vulnerability to and resilience of disease (Csete and Doyle, 1999, 2002; Csete and Doyle, 2004; Kitano, 2007; Doyle and Csete 2011).

(2) If the brain indeed constitutes a genuine networked system, i.e. non-trivial network structure plays an essential role in healthy brain function, then this structure could fundamentally affect or be affected by disease. Reciprocally, if network structure is essential to disease phenomenology, its characterisation could be an important step in the understanding of network structure's role in function. Moreover, if network structure is essential to both function and dysfunction, it could in principle be used for clinical purposes (Castellanos et al., 2013). For instance, it could define patient *theratypes* separating therapy responders from non-responders by encoding the structural mechanisms underlying disease expression and response to therapy (Agache and Akdis, 2019; Mobasheri and Loeser, 2024). Finally, if brain function and dysfunction become observable through network structure they could also be controllable through it. Thus, various fundamental questions concerning the role of complex network structure in disease, including whether there are genuine brain network diseases or whether network neuroscience can be used for disease classification (Douw et al., 2019), and more generally, for clinical purposes, hinge on the role of such a structure in brain function in general.

(3) The exact role of higher order network properties in disease aetiology and phenomenology is in general as yet insufficiently understood. This includes the appropriate space and scale in which a network structure may be meaningfully defined, the particular type of network property, as well as the way these may exert a role in brain function and dysfunction.

(4) At a theoretical level, it is still unclear whether there exist genuine network-related pathologies. Insofar as the role of complex network structure is disease-dependent, it may in general be more useful to think of networkness not as an all-or-nothing phenomenon, together with a set of necessary and sufficient conditions for its occurrence, but in terms of role of *network structure in disease* whose significance can only be determined by considering the brain function they support and the way they do it.

(5) At a practical level, current network theory's sensitivity and specificity is not (yet) sufficient to make it clinically relevant.

(6) The role of network geometry, topology and combinatorics in brain function and dysfunction is still insufficiently understood, and qualitative improvements require addressing the conceptual and methodological challenge represented by the translation of properties such as efficiency, homeostasis robustness, and evolvability in network terms that are also functionally meaningful when describing the brain (Levit-Binnun and Golland, 2012).

Morrell, F., Whisler, W.W., and Bleck, T.P. (1989). Multiple subpial transection: a new approach to the surgical treatment of focal epilepsy. *J. Neurosurg.* **70**, 231–239.

Mortimer, J.A. (1988). Do psychosocial risk factors contribute to Alzheimer's disease. In: A.S. Henderson, J.H. Henderson (Eds.), *Etiology of dementia of Alzheimer's type*, Wiley, New York, pp. 39–52.

Mueller, S.G., Laxer, K.D., Barakos, J., Cheong, I., Garcia, P., and Weiner, M.W. (2009). Widespread neocortical abnormalities in temporal lobe epilepsy with and without mesial sclerosis. *Neuroimage* **46**, 353–359.

Murphy, T.H., and Corbett, D. (2009). Plasticity during stroke recovery: from synapse to behaviour. *Nat. Rev. Neurosci.* **10**, 861–872.

Nelson, S.B., and Valakh, V. (2015). Excitatory/inhibitory balance and circuit homeostasis in autism spectrum disorders. *Neuron* **87**, 684–698.

Nicolle, C. (1933). *Leçons du Collège de France: destin des maladies infectieuses*. Paris: Felix Alcan.

Nijhout, H.F., Best, J.A., and Reed, M.C. (2019). Systems biology of robustness and homeostatic mechanisms. *Wiley Interdiscip. Rev.: Syst. Biol. Med.* **11**, e1440.

Nishimori, H., and Ortiz, G. (2010). *Elements of phase transitions and critical phenomena*. Oxford University Press, Oxford.

Nowotny, T., and Rabinovich, M.I. (2007). Dynamical origin of independent spiking and bursting activity in neural microcircuits. *Phys. Rev. Lett.* **98**, 128106.

Oberman, L., and Pascual-Leone, A. (2013). Changes in plasticity across the lifespan: cause of disease and target for intervention. *Prog. Brain Res.* **207**, 91–120.

Oberman, L.M, Rotenberg, A., and Pascual-Leone, A. (2014). Aberrant brain plasticity in autism spectrum disorders. In: *Cognitive plasticity in neurologic disorders*, Joseph I. Tracy, Benjamin M. Hampstead, and K. Sathian (Eds.), pp. 176–196.

Ocker, G.K., Josić, K., Shea-Brown, E., and Buice, M.A. (2017b). Linking structure and activity in nonlinear spiking networks. *PLoS Comput. Biol.* **13**, e1005583.

Ocker, G.K., Litwin-Kumar, A., and Doiron, B. (2015). Self-organization of microcircuits in networks of spiking neurons with plastic synapses. *PLoS Comput. Biol.* **11**, e1004458.

Ódor, G. (2008). *Universality in nonequilibrium lattice systems: theoretical foundations*. World Scientific.

Okano, H., Miyawaki, A., and Kasai, K. (2015). Brain/MINDS: brain-mapping project in Japan. *Phil. Trans. R. Soc. B* **370**, 20140310.

Olmi, S., Petkoski, S., Guye, M., Bartolomei, F., and Jirsa, V. (2019). Controlling seizure propagation in large-scale brain networks. *PLoS Comput. Biol.* **15**, e1006805.

Ostojic, S., and Fusi, S. (2024). Computational role of structure in neural activity and connectivity. *Trends Cognit. Sci.* **28**, 677.

Ott, E., Grebogi, C., and Yorke, J.A. (1990). Controlling chaos. *Phys. Rev. Lett.* **64**,1196.

Panas, D., Amin, H., Maccione, A., Muthmann, O., van Rossum, M., Berdondini, L., and Hennig, M.H. (2015). Sloppiness in spontaneously active neuronal networks. *J. Neurosci.* **35**, 8480-8492.

Pandya, S., Kuceyeski, A., and Raj, A. (2017). The brain's structural connectome mediates the relationship between regional neuroimaging biomarkers in Alzheimer's disease. *J. Alzheimers Dis.* **55**, 1639–1657.

Pandya, V.A., and Patani, R. (2021). Region-specific vulnerability in neurodegeneration: lessons from normal ageing. *Ageing Res. Rev.* **67**, 101311.

Papo, D. (2019a). Gauging functional brain activity: from distinguishability to accessibility. *Front. Physiol.* **10**, 509.

Papo, D. (2015). How can we study reasoning in the brain? *Front.Hum. Neurosci.* **9**, 222.

Papo, D. (2019b). Neurofeedback: principles, appraisal and outstanding issues. *Eur. J. Neurosci.* **49**, 1454–1469.

Papo, D. (2013b). Time scales in cognitive neuroscience. *Front. Physiol.* **4**, 86.

Papo, D., and Buldú, J.M. (2024). Does the brain behave like a (complex) network? I. Dynamics. *Phys. Life Rev.* **48**, 47–98.

Papo, D., and Buldú, J.M. (2025). Does the brain behave like a (complex) network? II. Function. (*in preparation*).

Papo, D., Zanin, M., Pineda, J.A., Boccaletti, S., and Buldú, J.M. (2014). Brain networks: great expectations, hard times, and the big leap forward. *Phil. Trans. R. Soc. B* **369**, 20130525.

Papo, I., Quattrini, A., Provinciali, L., Rychlicki, F., Del Pesce, M., Paggi, A., Ortenzi, F., Recchioni, M.A., and Censori, B. (1990). Callosotomy for the treatment of drug resistant generalized seizures. In *Neurosurgical Aspects of Epilepsy: Proceedings of the Fourth Advanced Seminar in Neurosurgical Research of the European Association of Neurosurgical Societies Bresseo di Teolo, Padova, May 17–18, 1989* (pp. 134–135). Springer Vienna.

Park, E., Velumian, A.A., and Fehlings, M.G. (2004). The role of excitotoxicity in secondary mechanisms of spinal cord injury: a review with an emphasis on the implications for white matter degeneration. *J. Neurotrauma* **21**, 754–774.

Parshani, R., Buldyrev, S.V., and Havlin, S. (2010). Interdependent networks: reducing the coupling strength leads to a change from a first to second order percolation transition. *Phys. Rev. Lett.* **105**, 048701.

Parshani, R., Rozenblat, C., Ietri, D., Ducruet, C., and Havlin, S. (2011). Inter-similarity between coupled networks. *EPL (Europhys. Lett.)* **92**, 68002.

Pascual-Leone, A., Freitas, C., Oberman, L., Horvath, J.C., Halko, M., Eldaief, M., Bashir, S., Vernet, M., Shafi, M., Westover, B., and Vahabzadeh-Hagh, A.M. (2011). Characterizing brain cortical plasticity and network dynamics across the age-span in health and disease with TMS-EEG and TMS-fMRI. *Brain Topogr.* **24**, 302–315.

Patania, A., Selvaggi, P., Veronese, M., Dipasquale, O., Expert, P., and Petri, G. (2019). Topological gene expression networks recapitulate brain anatomy and function. *Netw. Neurosci.* **3**, 744–762.

Pathak, A., Menon, S.N., and Sinha, S. (2024). A hierarchy index for networks in the brain reveals a complex entangled organizational structure. *Proc. Natl. Acad. Sci. U.S.A.* **121**, e2314291121.

Payne, J.L., and Wagner, A. (2019). The causes of evolvability and their evolution. *Nat. Rev. Genet.* **20**, 24–38.

Pazó, D., and Montbrió, E. (2006). Universal behavior in populations composed of excitable and self-oscillatory elements. *Phys. Rev. E* **73**, 055202.

Pecora, L.M., Sorrentino, F., Hagerstrom, A.M., Murphy, T.E., and Roy, R. (2014). Cluster synchronization and isolated desynchronization in complex networks with symmetries. *Nat. Commun.* **5**, 4079.

Penner, I.K., and Aktas, O. (2017). Functional reorganization is a maladaptive response to injury - NO. *Mult. Scler. J.* **23**, 193–194.

Peraza, L.R., Díaz-Parra, A., Kennion, O., Moratal, D., Taylor, J.P., Kaiser, M., Bauer, R., and Alzheimer's Disease Neuroimaging Initiative (2019). Structural connectivity centrality changes mark the path toward Alzheimer's disease. *Alzheimers Dement.* **11**, 98–107.

Pereira, J.B., Strandberg T.O., Palmqvist S., Volpe G., van Westen D., Westman E., Hansson O., and Alzheimer's Disease Neuroimaging Initiative (2018). Amyloid network topology characterizes the progression of Alzheimer's disease during the predementia stages. *Cereb. Cortex* **28**, 340–349.

Perl, Y.S., Fittipaldi, S., Campo, C.G., Moguilner, S., Cruzat, J., Fraile-Vazquez, M.E., Herzog, R., Kringelbach, M.L., Deco, G., Prado, P., and Ibanez, A. (2023). Model-based whole-brain perturbational landscape of neurodegenerative diseases. *eLife* **12**, e83970.

Pernice, V., Deger, M., Cardanobile, S., and Rotter, S. (2013). The relevance of network micro-structure for neural dynamics. *Front. Comput. Neurosci.* **7**, 72.

Perra, N., Baronchelli, A., Mocanu, D., Gonçalves, B., Pastor-Satorras, R., and Vespignani, A. (2012). Random walks and search in time-varying networks. *Phys. Rev. Lett.* **109**, 238701.

Peters, J.F. (2016). *Computational proximity*. Intelligent Systems Reference Library, vol. 102. Springer, Cham.

Peters, J.F. (2018). Proximal fiber bundles on nerve complexes. In: *Mathematical Analysis and Applications: Selected Topics*, (Ruzhansky, M., Dutta, H., & Agarwal, R. P. Eds.). pp. 517–536.

Petri, G., Musslick, S., Dey, B., Özcimder, K., Turner, D., Ahmed, N.K., Willke, T.L., and Cohen, J.D. (2021). Topological limits to the parallel processing capability of network architectures. *Nat. Phys.* **17**, 646–651.

Plis, S.M., Sui, J., Lane, T., Roy, S., Clark, V.P., Potluru, V.K., Huster, R.J., Michael, A., Sponheim, S.R., Weisend, M.P., and Calhoun, V.D. (2014). High-order interactions observed in multi-task intrinsic networks are dominant indicators of aberrant brain function in schizophrenia. *Neuroimage* **102**, 35–48.

Poo, M.M., Du, J.L., Ip, N.Y., Xiong, Z.Q., Xu, B., and Tan, T. (2016). China brain project: basic neuroscience, brain diseases, and brain-inspired computing. *Neuron* **92**, 591–596.

Porter, M.A., and Gleeson, J.P. (2016). Dynamical systems on networks. *Front. App. Dyn. Syst.* **4**, 1–80.

Pósfai, M., Szegedy, B., Bačić, I., Blagojević, L., Abért, M., Kertész, J., Lovász, L., and Barabási, A.L. (2024). Impact of physicality on network structure. *Nat. Phys.* **20**, 142–149.

Powell, F., Tosun, D., Sadeghi, R., Weiner, M., Raj, A., and Alzheimer's Disease Neuroimaging Initiative (2018). Preserved structural network organization mediates pathology spread in Alzheimer's disease spectrum despite loss of white matter tract integrity. *J. Alzheimers Dis.* **65**, 747–764.

Pozo, K., and Goda, Y. (2010). Unraveling mechanisms of homeostatic synaptic plasticity. *Neuron* **66**, 337–351.

Price, C.J., and Friston, K.J. (2002). Degeneracy and cognitive anatomy. *Trends Cognit. Sci.* **6**, 416–421.

Rabinovich, M., Huerta, R., and Laurent, G. (2008). Transient dynamics for neural processing. *Science* **321**, 48–50.

Radicchi, F., and Arenas, A. (2013). Abrupt transition in the structural formation of interconnected networks. *Nat. Phys.* **9**, 717–720.

Ramesh, P., Confavreux, B., Gonçalves, P.J., Vogels, T.P., and Macke, J.H. (2023). Indistinguishable network dynamics can emerge from unalike plasticity rules. *bioRxiv* 2023-11.

Raj, A., Kuceyeski, A., and Weiner, M. (2012). A network diffusion model of disease progression in dementia. *Neuron* **73**, 1204–1215.

Raj, A., and Powell, F. (2018). Models of network spread and network degeneration in brain disorders. *Biol Psychiatry Cogn. Neurosci. Neuroimaging* **3**, 788–797.

Rapisardi, G., Kryven, I., and Arenas, A. (2022). Percolation in networks with local homeostatic plasticity. *Nat. Commun.* **13**, 122.

Ratté, S., and Prescott, S.A. (2016). Afferent hyperexcitability in neuropathic pain and the inconvenient truth about its degeneracy. *Curr. Opin. Neurobiol.* **36**, 31–37.

Recanatesi, S., Pereira-Obilinovic, U., Murakami, M., Mainen, Z., and Mazzucato, L. (2022). Metastable attractors explain the variable timing of stable behavioral action sequences. *Neuron* **110**, 139–153.

Reis, S.D., Hu, Y., Babino, A., Andrade Jr, J.S., Canals, S., Sigman, M., and Makse, H.A. (2014). Avoiding catastrophic failure in correlated networks of networks. *Nat. Phys.***10**, 762–767.

Ren, X.L., Gleinig, N., Helbing, D., and Antulov-Fantulin, N. (2019). Generalized network dismantling. *Proc. Natl. Acad. Sci. U.S.A.* **116**, 6554–6559.

Riederer, F., Lanzenberger, R., Kaya, M., Prayer, D., Serles, W., and Baumgartner, C. (2008). Network atrophy in temporal lobe epilepsy: a voxel-based morphometry study. *Neurology* **71**, 419–425.

Rikkert, M.G.O., Dakos, V., Buchman, T.G., De Boer, R., Glass, L., Cramer, A.O., Levin, S., Van Nes, E., Sugihara, G., Ferrari, M.D., and Tolner, E.A. (2016). Slowing down of recovery as generic risk marker for acute severity transitions in chronic diseases. *Crit. Care Med.* **44**, 601–606.

Rings, T., Mazarei, M., Akhshi, A., Geier, C., Tabar, M.R.R., Lehnertz, K. (2019a). Traceability and dynamical resistance of precursor of extreme events. *Sci. Rep.* **9**, 1744.

Rings, T., von Wrede, R., and Lehnertz, K. (2019b). Precursors of seizures due to specific spatial-temporal modifications of evolving large-scale epileptic brain networks. *Sci. Rep.* **9**, 10623.

Roberts, J.A., Gollo, L.L., Abeysuriya, R.G., Roberts, G., Mitchell, P.B., Woolrich, M. W., and Breakspear, M. (2019). Metastable brain waves. *Nat. Commun.* **10**, 1056.
24